# Defects Vibrations Engineering for Enhancing Interfacial Thermal Transport in Polymer Composites

*Yijie Zhou, Robert Ciarla, Artittaya Boonkird, Thanh Nguyen, Jiawei Zhou, Zhang Jiang, Xiaobing Zuo, Jeewan Ranasinghe, Weiguo Hu, Brendan Scott, Shengxi Huang, Mingda Li, Yanfei Xu\**

Y. Zhou and B. Scott

Department of Mechanical and Industrial Engineering, University of Massachusetts Amherst, Amherst, Massachusetts, 01003, United States

R. Ciarla

Department of Chemical Engineering, University of Massachusetts Amherst, Amherst, Massachusetts, 01003, United States

A. Boonkird, T. Nguyen, and M. Li

Department of Nuclear Science and Engineering, Massachusetts Institute of Technology, Cambridge, Massachusetts, 02139, United States

J. Zhou

Department of Materials Science and Engineering, Stanford University, Stanford, California, 94305, United States

Z. Jiang and X. Zuo

Advanced Photon Source, Argonne National Laboratory, Argonne, Illinois, 60439, United States

J. Ranasinghe and S. Huang

Department of Electrical and Computer Engineering, Rice University, Houston, Texas, 77005, United States

W. Hu

Department of Polymer Science and Engineering, University of Massachusetts Amherst, Amherst, Massachusetts, 01003, United States

Y. Xu

Department of Mechanical and Industrial Engineering, University of Massachusetts Amherst; Department of Chemical Engineering, University of Massachusetts Amherst, Amherst, Massachusetts, 01003, United States

Email: yanfeixu@umass.edu

**Table of Contents**

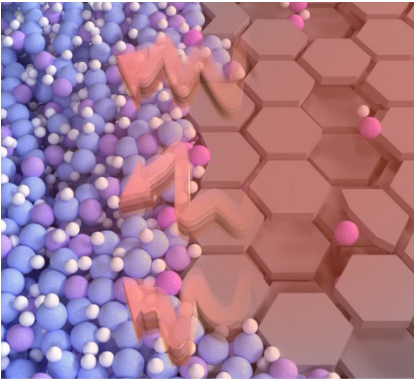

Defect engineering for improving heat conduction in polymer composites.

**Abstract**

To push upper boundaries of effective thermal conductivity in polymer composites, a fundamental understanding of thermal transport mechanisms is crucial. Although there is intensive simulation research, systematic experimental investigation on thermal transport in polymer composites is limited. To better understand thermal transport processes, we design polymer composites with "perfect" fillers (graphite) and defective fillers (graphite oxide); we choose polar polyvinyl alcohol (PVA) as a matrix model; and we identify how thermal transport occurs across heterogeneous interfaces. Measured thermal conductivities of ~1.44 $\pm$ 0.20 $W\ m^{-1}\ K^{-1}$ in PVA/defective filler composites is higher than those of ~0.87 $\pm$ 0.24 $W\ m^{-1}\ K^{-1}$ in PVA/"perfect" filler composites, while measured thermal conductivities in defective fillers are lower than those of "perfect" fillers. An effective quantum mechanical model is developed, showing that the vibrational state of the defective level plays a critical role in enhancing the thermal conductivity with increased defect concentration. Our experimental and model results have suggested that defects in polymer composites may enhance thermal transport in polymer composites by promoting vibrational resonant couplings.

**1. Introduction**

Polymer-based electronic devices, such as high-power batteries and soft robotics, have seen remarkable advancements in recent decades.[1-7] However, these devices often generate a significant amount of waste heat during operation, which can lead to overheating and safety hazards.[8, 9] Efficiently dissipating heat generated in polymer-based electronic devices is essential for ensuring their reliable and safe operation, but it remains a significant challenge.[10-12] Thermally conductive polymers are needed since thermally insulating polymers hinder heat dissipation.[13] Unfortunately, common polymers are thermal insulators with low thermal conductivities on the order of $0.1\ W\ m^{-1}\ K^{-1} - 0.3\ W\ m^{-1}\ K^{-1}$.[10, 14] To increase thermal conductivities ($k$) in polymers, highly thermally conductive fillers (e.g., carbon nanotube, graphene, and graphite, $k > 1000\ W\ m^{-1}\ K^{-1}$) have been added into polymers at high volume fractions (> 40 vol%).[15-21] However, measured thermal conductivity enhancement in these composites are generally limited to within one order of magnitude, which are much lower than theoretical predicted values.[22-24] Achieving polymer composites with enhanced thermal conductivity while using lower filler volume fractions would be a difficult yet extremely desirable goal.

Major challenges to achieving high effective thermal conductivities in polymer composites include high filler/filler and filler/polymer interfacial thermal resistances, a lack of effective control of the dispersion of thermally conductive fillers, and the requirement for filler loading at high volume fractions.[25, 26] High filler loadings in polymers deteriorate the mechanical performance and introduces processing challenges related to polymer's rheological behaviors.[27] Theoretical models, numerical simulations, and experimental studies have shown that improving electron transmission and/or phonon transmission are essential for enhancing interfacial thermal transport between two crystalline solids.[10, 28, 29] However, these interfacial thermal transport theories and simulations cannot be applied directly to polymers because polymer chains lack the regular periodicities and long-range orders that are characteristics of crystalline materials.[10, 30-32] It is known that an increase in interfacial defects can lead to a decrease in thermal conductance in various types of inorganic hybrid interfaces (e.g., Al/Si).[29, 33, 34] But there are exceptions.[29] Simulation results demonstrated that the thermal boundary conductance at interfaces between amorphous materials can be higher than that at interfaces between crystalline materials (e.g., Si/Ge).[35, 36] It remains difficult to predict and control interfacial thermal transport behaviors in polymer composites. The systematic experimental investigations of interfacial thermal transport mechanisms in polymer composites are limited compared to inorganic materials.[29, 37, 38] Understanding interfacial thermal transport mechanisms in polymer composites is challenging.[39] This is partially due to the complex structures in polymer composites, which are typically composed of multiple phases, disorders, and have high degrees of heterogeneities at multiple length scales, ranging from the molecular scale to the microscale.

In this study, to lay the foundation for understanding thermal transport mechanisms in polymer composites and controlling heat transfer across heterogenous interfaces, we address two fundamental questions: first, because defects are ubiquitous in polymer composites, can defects in fillers reduce filler/polymer interfacial thermal resistance and improve filler dispersion in polymers? Second, contradicting conventional understanding,[40, 41] can effective thermal conductivities in composites made of polymers and defective fillers with low thermal conductivities be higher than that of composites made of polymers and "perfect" fillers (graphite) with high thermal conductivities? To answer above questions, we design and synthesize "perfect" fillers (graphite) and defective fillers (graphite oxide) with controlled defects; we choose oxygen-containing polar polyvinyl alcohol (PVA) as a polymer matrix model for polymer composites (Figures 1). By having the interfacial oxygen-containing defects to couple with the polymers through the vibrational resonant couplings, we experimentally measured that high 418% increase of thermal conductivity is observed with as low as 5% volume fraction of fillers. We further develop an effective quantum mechanical model which could qualitatively explain the contradictory phenomena that we observe — increasing thermal conductivity despite decreasing heat capacity as defect density increases, which is possibly due to the vibrational levels of the defect states. This model further clarifies that composites consisting of polymers and defective fillers, characterized by low thermal conductivities, can demonstrate superior effective thermal conductivities compared to composites composed of polymers and "ideal" fillers such as graphite, which possess high thermal conductivities. Our experimental evidence confirms that defects can reduce interfacial thermal resistances, improve filler dispersions in polymers, and enhance effective thermal conductivities in polymer composites, supported by the effective theory. These could be done through the emergence of unique vibrational modes that arise from atomic defects at the filler/polymer interfaces with strong intermolecular noncovalent interactions (hydrogen bonds). This research may open exciting opportunities to design and create polymer composites as effective thermal interface materials with high thermal conductivities. Polymer-based thermal interface materials with high thermal conductivity are crucial components in various devices, including electronics, where they play a key role in transferring heat from devices to the environment.

## 2. Results and Discussion

To experimentally evaluate how defect vibrations enhance polymer/filler interfacial thermal transport through the emergence of unique vibrational modes intrinsic to the interfaces and defect atoms, we introduce oxygen-containing defects (oxygen and hydroxyl functional groups) on graphite surfaces and edges via graphite oxidation by a modified Hummers method (Figures 1C and 1D).[42-44] A statistical analysis of lateral sizes, thicknesses, and aspect ratios in "perfect" fillers (graphite) and defective fillers (graphite oxide) is in the supporting information (Figure S1).

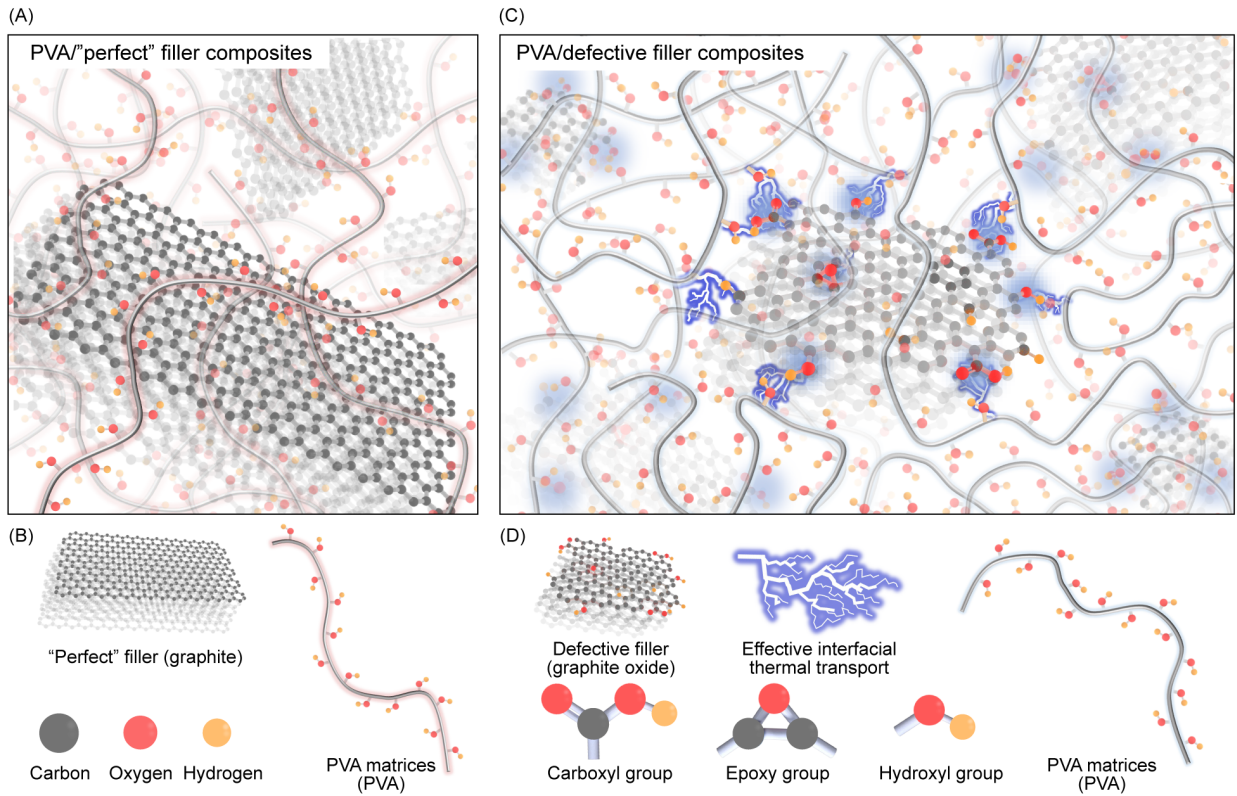

**Figure 1.** (A-B) Schematic illustration of PVA/"perfect" filler composites. The polymer matrix is oxygen-containing polar polyvinyl alcohol (PVA). The matrix is filled with high-purity graphite fillers, which are referred to as "perfect" fillers. (C-D) Schematic illustration of PVA/defective filler composites. The same polymer matrix made of oxygen-containing PVA. But it is filled with defective fillers (graphite oxide). To create these defective fillers, oxygen-containing defects (oxygen and hydroxyl functional groups) are introduced on the surfaces and edges of the graphite particles via graphite oxidation methods.[42-44] As a result, the surfaces and edges of defective graphite fillers becomes rough and uneven. When these defective graphite fillers are added to the polymer matrix, they create a heterogeneous composite material with heterogeneous interfaces. The defects in fillers may enhance interfacial thermal transport in polymer matrices.

To confirm the presence of defects in fillers, Raman spectra of defective and "perfect" graphite fillers are probed. A strong intensity of G band ~1582 $cm^{-1}$ that originates from the $E_{2g}$ vibration mode is observed in "perfect" fillers.[45-48] In contrast, a decreased intensity of G band ~1586 $cm^{-1}$ and an increased intensity of disorder-induced D band ~1350 $cm^{-1}$ are observed in defective fillers (Figure S2 in the supporting information).[45-48] This increased intensity of D band, related to the $A_{1g}$ breathing mode, is observed in defective fillers because oxidation of graphite alters the basal plane structure of graphite.[45-48] Defective graphite fillers have low in-plane thermal conductivities of ~63.93 ± 3.57 $W\ m^{-1}\ K^{-1}$ and low cross-plane thermal conductivities of

~1.87 ± 0.17 $W\ m^{-1}\ K^{-1}$ (Figures 2A and 2B). In contrast, "perfect" graphite fillers have high in-plane thermal conductivities of ~280.95 ± 13.54 $W\ m^{-1}\ K^{-1}$ and high cross-plane thermal conductivities of ~11.87 ± 0.96 $W\ m^{-1}$ (Figures 2A and 2B). The measured in-plane thermal conductivities in "perfect" graphite fillers are lower than those reported in a single crystal graphite.[19, 49] This might be due to the small crystallites in "perfect" graphite fillers and the presence of grain boundaries and defects.[19, 49] Measured values for specific heat capacities,[50] thermal diffusivities,[51] and densities in "perfect" fillers (graphite) and defective fillers (graphite oxide) are in the supporting information (Figure S3). Details for making pellet specimens for thermal diffusivity measurements in "perfect" fillers and defective fillers are in the supporting information (Section 1. experimental section).

To gain insight into thermal transport in polymer composites, thermal conductivities in polymer composites are determined by Equation 1. We probe the thermal diffusivity ($\alpha$) by a laser-flash technique and the specific heat capacity ($c_p$) by a differential scanning calorimetry technique. Density ($\rho$) is calculated by using mass divided by volume (Figure 2).

$$k = \alpha c_p \rho \quad (1)$$

Where $k$ is thermal conductivity ($W\ m^{-1}\ K^{-1}$), $\alpha$ is thermal diffusivity ($m^2\ s^{-1}$), $c_p$ is specific heat capacity ($J\ kg^{-1}\ K^{-1}$), and $\rho$ is density ($kg\ m^{-3}$).

Quantifying cross-plane thermal diffusivities and thermal conductivities is not only important for understanding thermal transport properties of polymer composites but also for optimizing performance for applications where efficient heat dissipation perpendicular to the surface is critical.[52, 53] The measured cross-plane thermal diffusivities in pure PVA films at room temperature are low, with a range of ~0.19 ± 0.02 $mm^2\ s^{-1}$ (Figure 2C). In contrast, the measured cross-plane thermal diffusivities in PVA/"perfect" filler composites (5 vol%) are higher, with a range of ~0.48 ± 0.13 $mm^2\ s^{-1}$. The cross-plane thermal diffusivities measured in polymer-defective filler composites (5 vol%) are even higher, with a range of ~0.83 ± 0.11 $mm^2\ s^{-1}$ (Figure 2C and Figures S4-S10 in the supporting information). Measured specific heat capacities in composites are relatively similar, with values ranging from ~1.32 ± $0.018\ J\ g^{-1}K^{-1}$ to ~$1.37\ \pm\ 0.016\ J\ g^{-1}K^{-1}$ (Figure 2D). The densities of PVA composites were measured (Figure 2E).

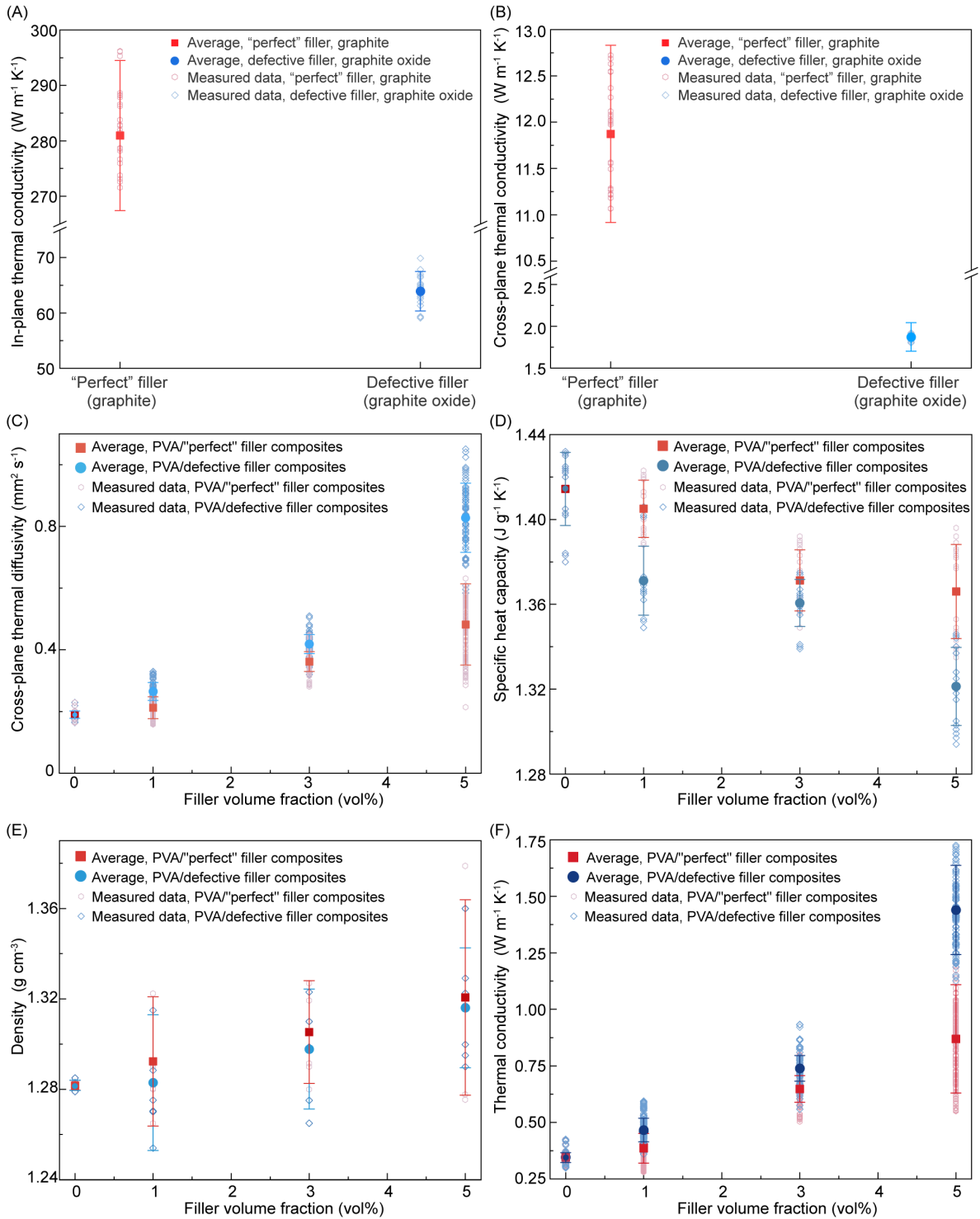

**Figure 2.** (A) Measured in-plane thermal conductivities of the "perfect fillers" (graphite) and the defective fillers (graphite oxide) at 25 °C. We performed three thermal diffusivity measurements for each sample, and we examined nine different samples from separate batches. (B) Measured cross-plane thermal conductivities of "perfect" fillers (graphite) and defective fillers (graphite oxide) measured at 25 °C. We performed three thermal diffusivity measurements for each sample, and we examined nine different samples from separate batches. The error bars of thermal conductivities of fillers are the error propagations based on cross-plane thermal diffusivity, specific heat capacity, and density measurements of fillers. Please refer to supporting information (Section 1) for detailed estimation of the error propagations of calculated thermal conductivity. (C) Measured cross-plane thermal diffusivities at 25 °C. The error bars of thermal diffusivity measurements mainly come from thickness differences of the polymer composites and thermal diffusivities differences among nine measurements for each sample. (D) Measured specific heat capacities at 25 °C. The error bars of specific heat capacities are the population standard deviation based on three measurements of each sample. (E) Measured densities at 25 °C. The error bars of the density are the population standard deviation based on six measurements of each sample. (F) Measured cross-plane thermal conductivities at 25 °C. The error bars of thermal conductivities are calculated based on error propagations of cross-plane thermal diffusivity, specific heat capacity, and density measurements of PVA composites. Please refer to the supporting information (Section 1) for detailed estimation of the error propagations of measured thermal conductivity.

Notably, measured cross-plane thermal conductivities in polymer/defective filler composites (5 vol%) and PVA/"perfect" filler composites (5 vol%) are ~$1.44 \pm 0.20\ W\ m^{-1}\ K^{-1}$ and ~$0.87 \pm 0.24\ W\ m^{-1}\ K^{-1}$, respectively (Figure 2F). However, measured thermal conductivities in defective fillers are lower than those of "perfect" fillers (Figures 2A and 2B). The X-ray scattering patterns indicate that there are no preferred polymer chain and filler orientations in these polymer composites (it will be discussed below in the Figure 4).[54] This could be favorable for achieving near-isotropic thermal conductivities in PVA composites. Thus, measured cross-plane thermal conductivities in polymer composites suggest the effective thermal conductivities in near-isotropic PVA-based composites (Figure 2F). Measured effective thermal conductivities with defective fillers are higher than those of PVA composites with "perfect" fillers with same filler loading ratios. The measured qualitative behaviors of the heat capacity and the thermal conductivity in Figures 2D and 2F can be explained through an effective quantum mechanical model developed in this work. The results of this model will be discussed in Figure 5 below and the supporting information (Section 3 and Figure S16).

From atomic level view of filler/polymer interfacial thermal transport, atoms in fillers interact with atoms in polymers via noncovalent interactions between PVA and fillers (graphite or graphite oxide). To provide insights into the factors that contribute to the highest thermal conductivities observed in these composites, we probe interfacial vibration couplings and dispersions of fillers in PVA using ATR-FTIR (attenuated total reflectance Fourier transform infrared spectroscopy) spectroscopy and $^{13}$C solid-state nuclear magnetic resonance (NMR) spectroscopy.

To experimentally confirm the presence of intermolecular vibration couplings arising from atomic mass defects at filler/polymer interfaces, we first examine vibration modes of functional groups on defective and "perfect" graphite fillers by using ATR-FTIR spectroscopy (Figure 3A). Stretching vibration peaks of oxygen-containing groups, including $1044\ cm^{-1}$ for alkoxy group (C-O), $1717\ cm^{-1}$ for carbonyl group (C=O), $3343\ cm^{-1}$ for hydroxyl group (-OH), are observed in the defective fillers.[55] Bending vibration peaks, including $1224\ cm^{-1}$ for epoxy groups (C-O-C) and $1363\ cm^{-1}$ for hydroxyl groups (C-OH), are observed in Figure 3A.[55] No vibration modes of oxygen-containing groups are observed on "perfect" graphite fillers (Figure 3A). Figure 3B shows ATR-FTIR spectra of PVA, PVA/"perfect" filler composites, and PVA/defective filler composites. The peak at $3276\ cm^{-1}$ in the ATR-FTIR spectra of PVA is attributed to O-H stretching vibration. This peak is broad, indicating that it arises from a range of different O-H bonds, including free hydroxyl groups and inter- or intramolecular hydrogen bonds.[56, 57] The addition of defective fillers to PVA matrices causes a shift in the strong vibration peak of PVA from $3271\ cm^{-1}$ to $3267\ cm^{-1}$ and $3266\ cm^{-1}$ with the addition of 1 vol% and 5 vol% defective fillers (graphite oxide), respectively (Figure 3B). These peak shifts suggest that there may be interactions between hydroxyl groups on PVA chains and functional groups on surfaces of defective fillers.[56, 58, 59] Interactions could be due to hydrogen bonding between the hydroxyl groups on PVA and the oxygen-containing functional groups on defective fillers, such as carbonyl, hydroxyl, and epoxy groups. This hydrogen bonding could result in a change in the vibrational energy of the hydroxyl groups on PVA, leading to the observed peak shifts. These hydrogen bonding interactions can cause a distortion of the electron cloud around the C-H and O-H bonds, resulting in a red-shift of the corresponding vibrational modes.[56, 58, 59] Similarly, interactions between hydroxyl groups on PVA chains and oxygen-containing functional groups on surfaces of defective fillers can cause a red-shift of the C-O bond peak at $1085\ cm^{-1}$. This red-shift is due to the increased electron density around the C-O bond because of hydrogen bonding interactions (Figure 3B).[58, 59] The presence of the "perfect" graphite fillers has not significantly affected molecular structures of PVA, since peaks observed in the PVA are still present in the composites (Figure 3B).[58, 59]

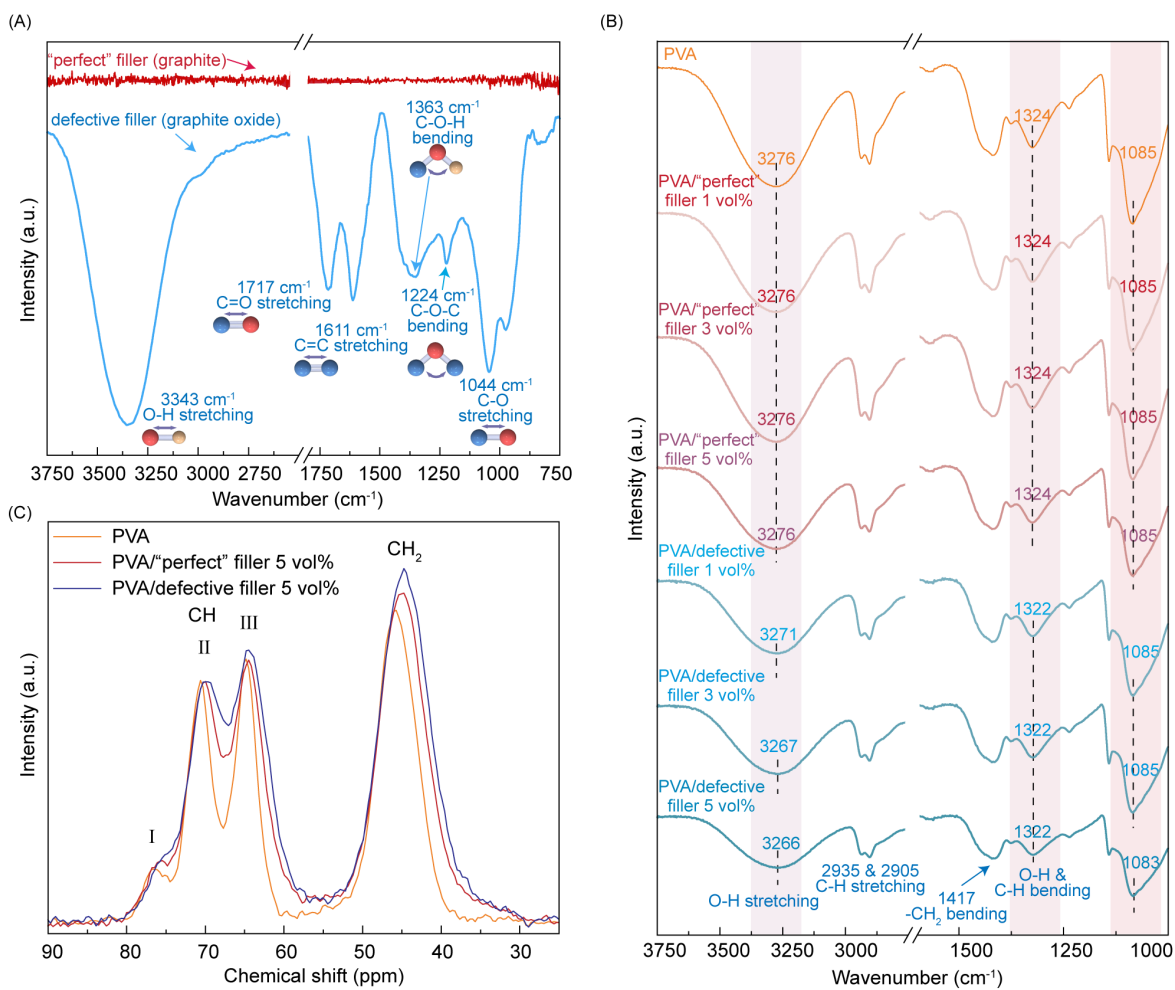

**Figure 3**. (A) The ATR-FTIR spectra of "perfect" fillers (graphite) and defective fillers (graphite oxide). (B) The ATR-FTIR spectra of PVA matrices, PVA/"perfect" filler composites, and PVA/ defective filler composites. (C) The NMR spectra of PVA matrices, PVA/"perfect" filler 5 vol% composites, and PVA/defective filler 5 vol% composites.

To experimentally confirm whether defects improve filler dispersions in polymers, we probe the structures of PVA chains and their interfaces with fillers using $^{13}C$ solid-state NMR spectroscopy at room temperature (Figure 3C).[60] Defective fillers (graphite oxide) and "perfect" fillers (graphite) are added into PVA matrices at 5 vol%. There is a tall peak at 46 ppm corresponding to $CH_2$ carbons and a group of three peaks between 60 and 80 ppm, conventionally named peak I, II, and III from downfield to upfield, corresponding to CH carbons whose OH forms 2, 1, and 0 hydrogen bonds with neighboring protons (Figure 3C).[61, 62] Signals from "perfect" fillers and defective fillers are undetectable. The chemical shift of the $CH_2$ signal in PVA observed in $^{13}C$ solid-state NMR spectrum is at 45.77 ppm. However, when PVA is incorporated with "perfect" fillers, the chemical shift of the $CH_2$ signal changes to ~45.07 ppm, when PVA is mixed with defective fillers, the chemical shift of the $CH_2$ signal shift changes to ~44.76 ppm. Similarly, peak II of the CH signal in the $^{13}C$ solid-state NMR spectroscopy of PVA is at 70.79 ppm, while it shifts to 70.27 ppm and 69.88 ppm in PVA/"perfect" filler composites and PVA/defective filler composites, respectively. These upfield (toward smaller ppm value) chemical shifts of $CH_2$ and CH signals could be due to two potential mechanisms: first, the presence of the conductive electrons in graphite or graphite oxide changes the magnetic field in the surrounding; and second, that the introduction of fillers into PVA matrices increases the population of *gauche* conformers in PVA chains.

In addition, the peak widths at half-maximum of the $CH_2$ signal in PVA, PVA/"perfect" filler composites, and PVA/defective filler composites are 891, 1062, 1116 Hz, respectively (Figure 3C). The $CH_2$ signal in PVA/defective filler composite is the broadest, while that in neat PVA is the narrowest. The CH signal shows similar peak broadening: peak II and peak III in neat PVA are resolved at around 40% peak maximum, while those in the "perfect" filler and defective filler PVA composites are resolved at around 70% and 80% maximum, respectively. The peak broadening in the PVA composites again could be due to two mechanisms: first, the presence of conductive electrons in the graphite or graphite oxide increase the magnetic field inhomogeneity in their surrounding; and second, the presence of the fillers broadens the distribution of the conformation of the PVA segments in their surroundings, which leads to a broadening of the NMR signals.[63-65] Both mechanisms of the influence of the fillers to the polymer matrix are short range, i.e., within a few nanometers. Therefore, larger peak position shifts and peak widths both indicate better dispersion of the fillers in the matrix.

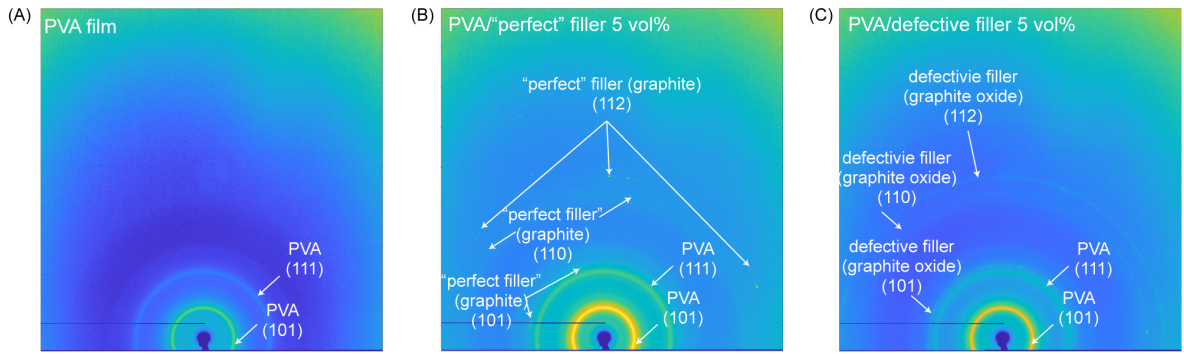

**Figure 4.** (A) Wide angle X-ray scattering (WAXS) patterns of PVA matrices. (B) WAXS patterns of PVA/"perfect" filler 5 vol% composites. (C) WAXS patterns of PVA/defective filler 5 vol% composites.

To gain a better understanding of relationships between thermal transport and structures in polymers and polymer composites, we utilized the synchrotron wide-angle X-ray scattering technique to investigate orientations of fillers and polymer chains (Figure 4 and Figure S15).[66] Our observations reveal characteristic Bragg scatterings, which include the (101) plane groups in PVA matrix and (101) plane groups in graphite-based fillers (Figures 4B and 4C).[67, 68] The X-ray scattering pattern also displayed multiple concentric rings, which are characteristic of Bragg scattering off different planes of crystals. The presence of these concentric rings rather than distinct dots in patterns suggests that there are no preferred orientations of crystallites within polymer composites.[66, 69] The measured cross-plane thermal conductivities in polymer composites suggest those in near-isotropic PVA-based composites (Figure 2F).

We observed intriguing thermal transport phenomena: First, the specific heat capacity decreased while the thermal conductivities of polymer composites increased as the filler fraction ratio increased (Figures 2D and 2F). Second, effective thermal conductivities in composites made of polymers and defective fillers with low thermal conductivities can be higher than those of composites made of polymers and "perfect" fillers (graphite) with high thermal conductivities (Figures 2A, 2B, and 2F). To better understand these unusual thermal transport phenomena, an effective quantum mechanical model is developed from the coupling of the vibrational state of the backbone polymer chain and the fillers (Section 3 in the supporting information). The simple Hamiltonian of such system can be written as Equation 2:

$$H = \sum_{\mathbf{q}} \omega_{\mathbf{q}} \left(a_{\mathbf{q}}^+ a_{\mathbf{q}} + \frac{1}{2}\right) + \omega_1 \left(a_1^+ a_1 + \frac{1}{2}\right) + \omega_2 \left(a_2^+ a_2 + \frac{1}{2}\right) +$$

$$\sum_{\mathbf{q}j} (V_{\mathbf{q}j} a_{\mathbf{q}}^+ a_j + V_{\mathbf{q}j}^* a_j^+ a_{\mathbf{q}}) + V_{12} a_1^+ a_2 + V_{12}^* a_2^+ a_1 \quad (2)$$

Where $\omega_1$, $\omega_{\mathbf{q}}$, and $V_{\mathbf{q}j}$ denote the phonon vibrational energy of the fillers, the polymer backbone, and the coupling potential between fillers and polymer backbone, respectively. The standard Bosonic communications relation hold for both polymer backbone and fillers, i.e., $[a_{\mathbf{q}}, a_{\mathbf{k}}^+] = \delta_{\mathbf{qk}}$, $[a_i, a_j^+] = \delta_{ij}$ etc. For defective fillers, additional energy level $\omega_2$ is introduced, and $V_{12}$ denotes the tunneling between the two defect modes. This model is inspired by Anderson's two-level system model of thermal transport in disordered solids, where the vibrational coupling becomes dominant,[70] and has been used to explain complex thermal transport process in interacting defects systems.[71] The closed form solution of self-energy, group velocity, and thermal conductivity can be derived from Green's function equation of motion approach. The numerical result of the effective quantum mechanical model is shown in Figure 5. The details for specific heat capacity calculation are in the supporting information Equation S3.4.5. In region II and III (Figure 5A), the decreasing of the specific heat capacity $C(T)$ with increasing filler fraction ($n_i$) links to the decreasing of the renormalized vibrational energy ($\omega_k'$) (Figures 5C and 5D). The increasing of thermal conductivity $K(T)$ in region II is a result of higher increasing of phonon group velocity ($v_k$) compared to the phonon scattering ($\tau_k$). This trend generally occurs when the coupling potential $V_k$ is weaker than $\omega_k$, while $\omega_1$ may be smaller or comparable to $\omega_k$. On the other hand, if the increasing of phonon scattering dominates the enhancement of the group velocity, the thermal conductivity will decrease with $n_i$ (region III).

It is found that measured thermal transport properties in composites including PVA/"perfect" graphite filler composites and PVA/defective filler graphite oxide composites (Figures 2D and 2F) located in modelled region II (Figures 5A and 5C), where the increased fraction of filler leads to the increased thermal conductivities and decreased specific heat capacity of the composites. Based on the effective quantum mechanical model, it is anticipated that with the increased fraction of fillers, increase in thermal conductivity of the polymer composite can be observed due to increased phonon group velocity in region II (Figure 5A).[72]

Finally, for defective fillers, which can be modeled as non-degenerate defects with $\omega_1 \neq \omega_2$, will lead to a higher thermal conductivity $K(T)$ but lower heat capacity $C(T)$ compared to the perfect filler (modelled as $\omega_1 = \omega_2$), as discussed in Section 3.5 in the supporting information. These effective quantum mechanical model results agree with experimental observations: composites made of polymers and defective fillers with low thermal conductivities can exhibit higher effective thermal conductivities than those in composites made of polymers and "perfect" fillers like graphite, which have high thermal conductivities. Coupling with more defect channels can lead to an additional hardening of the backbone of polymers, leading to the above observations. The coarse-grain potentials in PVA-PVA, PVA-defective filler (graphite oxide), and defective filler- defective filler was obtained in the supporting information Figure S16 .[72]

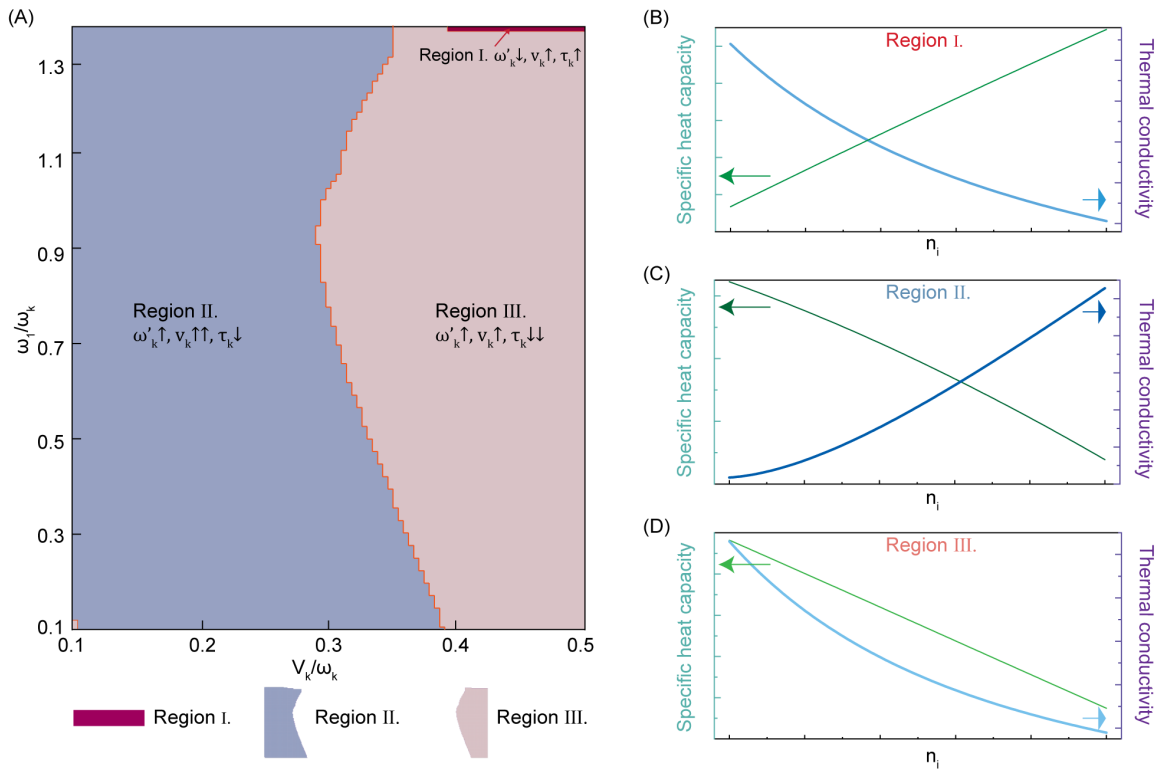

**Figure 5.** (A) The numerical results of the coupled vibration state model of polymer composite. The results are correlated to the vibrational energy of the fillers ($\omega_1$), vibrational energy of the polymer backbone ($\omega_k$), and the coupling potential between fillers and polymer backbone ($V_k$). The region I, II, and III indicates the different behavior of thermal conductivity $K(T)$ and specific heat capacity $C(T)$ with the increasing filler fraction ($n_i$) of the polymer composites. The double arrows indicate the domaining effect in specific region. The renormalized phonon energy of polymer backbone ($\omega'_k$), group velocity ($v_k$) and phonon relaxation time ($\tau_k$) are interpreted for explaining the behavior of the changing $C(T)$ and $K(T)$. (B) In region I, with the increasing filler fraction, $C(T)$ increases while $K(T)$ decreases. (C) In region II, the increasing filler fraction

results in decreased $C(T)$ and increased $K(T)$, which matches the results observed in PVA/defective filler composites. (D) In region III, both $C(T)$ and $K(T)$ decreases as the increasing of filler fraction.

## 3. Conclusion

This work presents experimental evidence that challenges traditional understandings of interfacial defects as adding resistance to heat transfer across heterogenous interfaces. Instead, we demonstrate that interfacial defects can enhance effective thermal conductivities in polymer composites through vibrational couplings that arise from oxygen-containing defects. Specifically, measured cross-plane thermal conductivities ~$1.44 \pm 0.20\ W\ m^{-1}\ K^{-1}$ in PVA/defective filler composites (graphite oxide, 5 vol%) are higher than those ~$0.87 \pm 0.24\ W\ m^{-1}\ K^{-1}$ in PVA/"perfect" filler composites (graphite, 5 vol%). However, measured in-plane thermal conductivities in defective fillers (~$63.93 \pm 3.57\ W\ m^{-1}\ K^{-1}$) are lower than those of "perfect" fillers (~$280.95 \pm 13.54\ W\ m^{-1}\ K^{-1}$). Interfacial defects enhance thermal transport across heterogenous interfaces in polymer composites. Our numerical result well explained the observed opposite trend between decreasing specific heat capacity and increasing thermal conductivity in polymer composites.

By understanding and controlling interfacial thermal transport, it is possible to reduce interfacial thermal resistances in polymer composites with high densities of heterogenous interfaces. Through manipulating the local heat flux carried by atomic vibrations that comprise these interfacial vibration couplings, effective thermal conductivities of polymer composites can be enhanced. By developing polymer composites with high cross-plane thermal conductivities, it is possible to create effective thermal interface materials that can be used in various sustainable applications such as thermal management of electronic devices, aerospace materials, and energy storage.

## 4. Experimental Section

**Materials.** All chemicals were purchased and used as received. Polyvinyl alcohol (PVA, $M_w$ 89000-98000, 99+% hydrolyzed, CAS: 9002-89-5), potassium permanganate (CAS: 7722-64-7), and graphite (flakes, ≥98% carbon basis, +50 mesh particle size (≥80%), natural, CAS: 7782-42-5) were purchased from Sigma-Aldrich. Concentrated sulfuric acid (CAS: 7664-93-9) was purchased from Alfa Aesar. Sodium nitrate (CAS: 7631-99-4), hydrogen peroxide (CAS: 7722-84-1), and ethanol (CAS: 64-17-5) were purchased from Thermo Fisher Scientific.

**Preparation of defective fillers (graphite oxide).** Defective fillers (graphite oxide) with oxygen-containing defects (oxygen and hydroxyl functional groups) on graphite surfaces and edges were prepared via the modified Hummers method.[42-44] Typically, graphite flakes (3.0 g) were mixed with concentrated sulfuric acid (69.0 mL) and sodium nitrate (1.5 g) in a 500 mL flask at 0 °C with continuous stirring for 30 minutes. After that, potassium permanganate (9.0 g) was added into a reaction mixture in small portions to prevent the temperature from rising above 10 °C. Then the temperature of the reaction mixture was raised to 35 °C using an oil bath, the mixture was stirred for 30 minutes. After completion of the reaction, 138 mL of deionized water was gradually added into the reaction. Then the temperature of the reaction mixture was raised to ~98 °C using an oil bath, and the mixture was stirred for 15 minutes. The suspension was reacted further by adding a mixture of hydrogen peroxide (6 mL) and water (23 mL). The mixture was stirred for 1 hour. After the graphite oxide was formed through the procedure described above, it was separated from the suspension using a centrifuge. The centrifugation was carried out at a speed of 10000 rpm for 10 minutes. The separated graphite oxide was washed with water using an ultrasonic bath for 8 hours. The next step involved centrifuging graphite oxide multiple times and washing it with water until the pH of the supernatant (the liquid above the solid material) reached 7. After the purifying process, graphite oxide was dried in an oven at a temperature of 110 °C for 24 hours before it was further used in the experiment.

**Preparation of PVA solution and PVA thin films.** A solution containing 8 wt% of PVA was prepared using deionized water as the solvent. Typically, PVA powders (8 g) were added into deionized water (92 g). Then, the temperature of the mixture was raised to ~ 85 °C using an oil bath, and the mixture was stirred for 3 hours until the PVA was dissolved. The solution was then allowed to cool to room temperature and was stirred for an additional hour. The PVA films (thickness ~20 μm) for structural and thermal transport property studies were prepared by casting PVA solution using a coating rod (Buschman HS75) at room temperature. The PVA films were dried at 95 °C for 15 minutes by a hot plate to remove the solvent. PVA films were finally dried in an oven at a temperature of 95 °C for 12 hours before they were further used in the experiment.

**Preparation of PVA/defective filler composite films or PVA/"perfect" filler composite films.** The procedure involved dispersing the fillers (graphite oxide or graphite) in PVA solution using ultrasonication and constant stirring to obtain a homogeneous dispersion. This step was repeated three times to ensure complete dispersion of the fillers in the PVA solution. Once the dispersion was prepared, it was cast onto a clean substrate using a coating rod (Buschman HS75) to achieve a uniform film under room temperature. The film was then dried using a hot plate at 95 °C for 5 minutes to remove the excess water and further dried in an oven at 95 °C for 12 hours to ensure complete removal of the water and to form a solid composite film. The "perfect" filler composite preparation follows the same procedure as the defective filler composite preparation, except for the type of filler material used. For example, in order to make composite films made of PVA/defective filler (5 vol%) composites, graphite oxide powders (0.033 g) were added into PVA solution (8 wt%, 5 mL). This solution was dispersed using an ultrasonic bath for 30 minutes. After ultrasound treatment, the dispersion was followed up with a constant stirring for 8 hours. Such procedure of ultrasonication and constant stirring was repeated three times in total. These steps were necessary to ensure that fillers were dispersed in the PVA solution. The dispersion was then cast on a clean glass substrate. A coating rod (Buschman HS75) was used to make a uniform film. The film was first dried on a hot plate at 95 °C for 5 minutes and then dried at 95 °C for 12 hours in an oven.

**Supporting Information**

General details of experimental sections (characterizations), supporting data for thermal diffusivities and specific heat capacities, one-dimensional curves of synchrotron X-ray scattering intensities versus scattering vectors, and effective theory of dynamical defects enhanced lattice thermal conductivity.

**Conflicts of interest**

There are no conflicts to declare.

**Acknowledgements**

Yijie Zhou, Robert Ciarla, Brendan Scott, and Yanfei Xu acknowledge the Faculty Startup Fund support from the University of Massachusetts Amherst and the support from National Science Foundation (award number 2312559). They also thank Dhandapani Venkataraman from the University of Massachusetts Amherst for allowing them to access attenuated total reflectance Fourier transform infrared spectroscopy. Jeewan Ranasinghe and Shengxi Huang acknowledge the support from National Science Foundation with grant numbers ECCS-1934977 and ECCS-2246564. Jiang Zhang and Xiaobing Zuo acknowledge support from Advanced Photon Source, a U.S. Department of Energy (DOE) Office of Science user facility operated for the DOE Office of Science by Argonne National Laboratory under Contract No. DE-AC02-06CH11357. Mingda Li acknowledges support from National Science Foundation (NSF) Designing Materials to Revolutionize and Engineer our Future (DMREF) Program with award number DMR-2118448. Thanh Nguyen is supported by the NSF Convergence Accelerator number No. 2235945.

**Supporting Information for**

**Defects Vibrations Engineering for Enhancing Interfacial Thermal Transport in Polymer Composites**

Yijie Zhou,[1] Robert Ciarla,[2] Artittaya Boonkird,[3] Thanh Nguyen,[3] Jiawei Zhou,[4] Zhang Jiang,[5] Xiaobing Zuo,[5] Jeewan Ranasinghe,[6] Weiguo Hu,[7] Brendan Scott,[1] Shengxi Huang,[6] Mingda Li,[3] and Yanfei Xu[1,2,*]

[1] Department of Mechanical and Industrial Engineering, University of Massachusetts Amherst, Amherst, Massachusetts, 01003, United States

[2] Department of Chemical Engineering, University of Massachusetts Amherst, Amherst, Massachusetts 01003, United States

[3] Department of Nuclear Science and Engineering, Massachusetts Institute of Technology, Cambridge, Massachusetts 02139, United States

[4] Department of Materials Science and Engineering, Stanford University, Stanford, California 94305, United States

[5] Advanced Photon Source, Argonne National Laboratory, Argonne, Illinois 60439, United States

[6] Department of Electrical and Computer Engineering, Rice University, Houston, Texas 77005, United States

[7] Department of Polymer Science and Engineering, University of Massachusetts Amherst, Amherst, Massachusetts 01003, United States

* Correspondence to: yanfeixu@umass.edu

**Contents**

**Section 1. Experimental section**

1.1 Characterization.

1.2 Calculation of population standard deviation and error propagation.

**Section 2. Supporting data**

Figure S1. Statistical analyses of lateral sizes, thickness, and aspect ratios in "perfect" fillers (graphite) and defective fillers (graphite oxide).

Figure S2. Raman spectra of the "perfect" fillers (graphite) and the defective fillers (graphite oxide).

Figure S3. In-plane and cross-plane thermal diffusivities, specific heat capacities, and densities of compressed "perfect" fillers (graphite) and defective fillers (graphite oxide) measured at 25 °C.

Figure S4. Cross-plane thermal diffusivities of PVA films measured at 25 °C.

Figure S5. Cross-plane thermal diffusivities of PVA/defective filler 1 vol% composites films measured at 25 °C.

Figure S6. Cross-plane thermal diffusivities of PVA/defective filler 3 vol% composite films measured at 25 °C.

Figure S7. Cross-plane thermal diffusivities of PVA/defective filler 5 vol% composite films measured at 25 °C.

Figure S8. Cross-plane thermal diffusivities of PVA/"perfect" filler 1 vol% composite films measured at 25 °C.

Figure S9. Cross-plane thermal diffusivities of PVA/"perfect" filler 3 vol% composite films measured at 25 °C.

Figure S10. Cross-plane thermal diffusivities of PVA/"perfect" filler 5 vol% composite films measured at 25 °C.

Figure S11. Cross-plane thermal diffusivities of compressed graphite oxide measured at 25 °C.

Figure S12. In-plane thermal diffusivities of compressed graphite oxide measured at 25 °C.

Figure S13. Cross-plane thermal diffusivities of compressed graphite measured at 25 °C.

Figure S14. In-plane thermal diffusivities of compressed graphite measured at 25 °C.

Figure S15. One-dimensional curves of synchrotron X-ray scattering intensities versus scattering vectors ($q$) for polymers and polymer composites: compressed graphite, compressed graphite

oxide, PVA film. PVA/"perfect" filler 5 vol % composite films, and PVA/defective filler 5 vol % composite films.

Figure S16. Relationships between coarse-grain potential to bond lengths in various materials, including polymer-polymer (PVA-PVA), polymer-filler (PVA-graphite oxide), and filler-filler (graphite oxide-graphite oxide).

**Section 3. Effective theory of dynamical defects enhanced lattice thermal conductivity.**

3.1 Effective quantum mechanical model (vibrational Hamiltonian model) setup.

3.2 Green's functions.

3.3 Phonon Green's function in polymer matrix.

3.4 Kubo formula for lattice thermal conductivity of defective polymers.

3.5 The connection of experimental data.

3.6 Numerical analysis.

## Section 1. Experimental section.

### 1.1 Characterization.

**ATR-FTIR (Attenuated total reflectance Fourier transform infrared spectroscopy) analysis.** The ATR-FTIR was carried out on a Bruker ALPHA II FTIR system. The ATR-FTIR measurements were performed by scanning the spectrum from 4000 $cm^{-1}$ to 400 $cm^{-1}$, with a resolution of 2 $cm^{-1}$. Each sample was scanned 16 times to improve the signal-to noise ratio, and the average value was taken. Before the ATR-FTIR measurements, samples were dried using an oven at 110 °C for 12 hours to remove any moisture. The ATR-FTIR measurements were taken at room temperature under ambient atmosphere. Measured data were corrected using automatic baseline corrections.

**Raman spectroscopy analysis.** The Raman spectra were recorded on a Horiba LabRAM Raman spectrometer. The Raman spectrometer was equipped with a 600/mm grating and a 50x objective. The acquisition time was 30 s. The laser power on the sample was kept at 10 mW. The excitation laser wavelength was 532.5 nm. In each Raman measurement, two accumulations were carried out. Thus, each spectrum was averaged three measurements. Raman measurements were taken at room temperature under ambient atmosphere.

**$^{13}C$ solid-state nuclear magnetic resonance (NMR) analysis.** $^{13}C$ cross polarization with magic angle spinning (CP/MAS) experiments were performed on a Bruker 600 MHz solid-state NMR spectrometer in a 4 mm broadband-observe CP/MAS probe. The spinning speed was 8 kHz. A contact time of 2 ms, a recycle delay of 20 s, and a decoupling field strength of 70 kHz were used for NMR experiments. Chemical shift was calibrated by setting the unprotonated aromatic carbon signal of 1,4-di(*t*-butyl)benzene at 148.8 ppm.

**Synchrotron X-ray scattering measurement.** Synchrotron X-ray scattering experiments were conducted at beamline 12-ID-B of the Advanced Photon Source located at Argonne National Laboratory. The wavelength λ of the X-ray beam is 0.93 Å (13.3 keV). The sample-detector distance was 188.389 mm. Thin polymer composite films were folded to achieve a thickness above 0.25 mm for synchrotron X-ray scattering experiments. To make graphite and graphite oxide flakes pallet specimens for synchrotron X-ray scattering measurements, the amount of flakes being used was between 0.15 to 0.45 grams, and they were poured into a customized mold with a diameter of 25.4 millimeters. Graphite was purchased from Sigma-Aldrich (size of 50 mesh, CAS: 7782-42-5); graphite oxide was prepared via the modified Hummers method.[1-3] A hydraulic press was used to apply pressure to the graphite or graphite oxide flakes that were poured into the customized mold, resulting in the formation of pellet specimens. Typically, after applying a pressure of 58 MPa using a hydraulic press for 5 minutes, the pellet specimen was removed from the mold.

**Thermal diffusivity measurements.** Cross-plane thermal diffusivities in PVA films and PVA composites were measured by laser flash apparatus with a lamp voltage of 150 V and a pulse width of 50 μs (LFA 467 HyperFlash, NETZSCH). All PVA films and polymer composites were sprayed with graphite (DGF 123, Miracle Power Products) prior to measuring cross-plane thermal diffusivities. A transparent model provided by the NETZSCH software was used for analyzing the data obtained from laser flash experiments and extracting thermal diffusivity. The densities of compressed graphite or graphite oxide flakes can be calculated by measuring the thickness and weight of a pellet specimen made from the flakes. The in-plane and cross-plane thermal diffusivities of the compressed flakes are measured using a lamp voltage of 250 V and a pulse width of 600 μs. The penetration model provided by the NETZSCH software was used for analyzing the data obtained from laser flash experiments and extracting the cross-plane thermal diffusivities of compressed graphite and graphite oxide. The in-plane anisotropic model provided by the NETZSCH software was used for extracting the in-plane thermal diffusivities of compressed graphite and graphite oxide.

**Preparation of pellet specimens for thermal diffusivity measurements in "perfect" fillers (graphite) and defective fillers (graphite oxide).** To measure the cross-plane and in-plane thermal diffusivities of graphite purchased from Sigma-Aldrich (flakes, ≥98% carbon basis, +50 mesh particle size (≥80%), natural, CAS: 7782-42-5) and graphite oxide made by modified Hummers method, samples were prepared using the following procedures. Dried graphite or graphite oxide fillers (0.25 g) were first dispersed in deionized water (50 mL) to make 5 g/L dispersion. The dispersion was then ultrasonicated for 30 minutes using an ultrasonic cleaner (VWR Symphony 97043-936), followed by constant stirring for 8 hours at room temperature. This process of ultrasonication (30 minutes) and stirring (8 hours) was repeated three times in total. After the ultrasonication and stirring process, the dispersion was dried in an oven (Thermo Scientific Blue M vacuum oven VO914A1) at 110 °C for 24 hours. After drying, the ultrasound-treated graphite fillers or graphite oxide fillers were weighed out (around 0.15 to 0.45 g) as per experimental design and were poured into a customized mold with a diameter of 25.4 mm, to make a filler pellet specimen for thermal diffusivity testing. A pressure of 58 MPa was applied to the

mold using a hydraulic press for 5 minutes to form the pellet specimen. For the in-plane thermal diffusivity tests, the pellet specimen was carefully removed from the mold. For the cross-plane thermal diffusivity tests, the pellet specimen was further cut into a smaller pellet with a diameter of 12.7 mm using a hollow punch with a diameter of 12.7 mm.

**Preparation of polymer films and polymer composite films for thermal diffusivity measurements.** To measure the cross-plane thermal diffusivities, PVA films, PVA/defective filler composite films, and PVA/"perfect" filler composite films with thicknesses of ~15-30 μm were cut into a round-shape films (12.7 mm in diameter) by using a hollow punch with a diameter of 12.7 mm. The films were first sprayed with graphite coatings (DGF 123, Miracle Power Products) on both sides and were then kept at room temperature for 5 minutes until the graphite spray dried. The details for making PVA films, PVA/defective filler composite films, and PVA/"perfect" filler composite films were in the main text (experimental section)

**Specific heat capacity measurements.** The specific heat capacities of samples were measured using differential scanning calorimetry (DSC 2500, TA Instruments). In DSC measurements, samples were heated at a ramp rate of 10 °C/minute until it reached a temperature of 130 °C. Then it was cooled down to 0 °C at the same ramp rate of 10 °C/minute. DSC measurements involved repeating the heating and cooling cycle four times, with the first cycle used to eliminate any thermal history of the sample and the last three cycles used to obtain the specific heat capacity of the sample. Sample weight in DSC experiments was in the range of 8 to 12 mg. During the DSC measurement, nitrogen was used as a DSC standard purge gas and was kept purging with a gas flow of 400 mL/ minute.

**Lateral size and thickness characterization of graphite and graphite oxide fillers.** The lateral sizes and thicknesses of the graphite and graphite oxide fillers were measured by Zygo Nexview 3D optical profiler (Nexview, Metek). Before lateral sizes and thicknesses measurements using Zygo system, the fillers were first dispersed in deionized water to make a 1 g/L dispersion. The dispersion was then dropped on a clean silica substrate and heated on a hot plate at 95 °C for 10 minutes. This step was done to evaporate the water and ensure that the fillers were evenly distributed on the substrates.

**Scanning electron microscope measurements.** The SEM images of the fillers were taken by FEI Magellan 400 XHR Scanning Electron Microscope.

## 1.2 Population standard deviation and error propagation.

The population standard deviation ($\sigma$) is determined by Equation S1.2.1.

$$\sigma = \sqrt{\frac{\Sigma(x_i-\mu)^2}{N}} \quad \text{(Equation S1.2.1)}$$

Where $\sigma$ is the population standard deviation, $\Sigma$ is the sum from 1 to N, N is the total number of the population, $x_i$ is an individual value (e.g., thermal diffusivity, specific heat capacity, or density), and $\mu$ is the population mean.

The population mean ($\mu$) is determined by Equation S1.2.2.

$$\mu = \bar{x} = \frac{1}{N}\left(\sum_{i=1}^{N} x_i\right) = \frac{x_1 + x_2 + \cdots + x_N}{N} \quad \text{(Equation S1.2.2)}$$

Where $\mu$ (or $\bar{x}$) is the population mean, $\Sigma$ is the sum from 1 to N, N is the total number of the population, and $x_i$ is an individual value (e.g., thermal diffusivity, specific heat capacity, or density).

Thermal conductivities ($k$) in polymer composites are determined by Equation S1.2.3 (the same as Equation 1 in the main text)

$$k = \alpha C_p \rho \quad \text{(Equation S1.2.3)}$$

Where $k$ is thermal conductivity ($W\ m^{-1}\ K^{-1}$), $\alpha$ is thermal diffusivity ($m^2\ s^{-1}$), $C_p$ is specific heat capacity ($J\ kg^{-1}\ K^{-1}$), and $\rho$ is density ($kg\ m^{-3}$).

The error propagation of $k$ is determined by Equations S1.2.4 and S1.2.5.

$$\Delta k = \sqrt{\left(\frac{\partial k}{\partial \alpha}\right)^2 (\sigma_\alpha)^2 + \left(\frac{\partial k}{C_p}\right)^2 (\sigma_{C_p})^2 + \left(\frac{\partial \rho}{\rho}\right)^2 (\sigma_\rho)^2} \quad \text{(Equation S1.2.4)}$$

which equals to

$$\Delta k = \sqrt{\left(\overline{C_p} \times \bar{\rho}\right)^2 (\sigma_\alpha)^2 + (\bar{\alpha} \times \bar{\rho})^2 (\sigma_{C_p})^2 + \left(\bar{\alpha} \times \overline{C_p}\right)^2 (\sigma_\rho)^2} \quad \text{(Equation S1.2.5)}$$

Where $\Delta k$ is the error propagation of thermal conductivity ($W\ m^{-1}\ K^{-1}$), $\sigma_\alpha$ is the population standard deviation of thermal diffusivity ($m^2\ s^{-1}$), $\sigma_{C_p}$ is the population standard deviation of the specific heat capacity ($J\ kg^{-1}\ K^{-1}$), and $\sigma_\rho$ is the population standard deviation of density ($kg\ m^{-3}$).

**Section 2. Supporting data.**

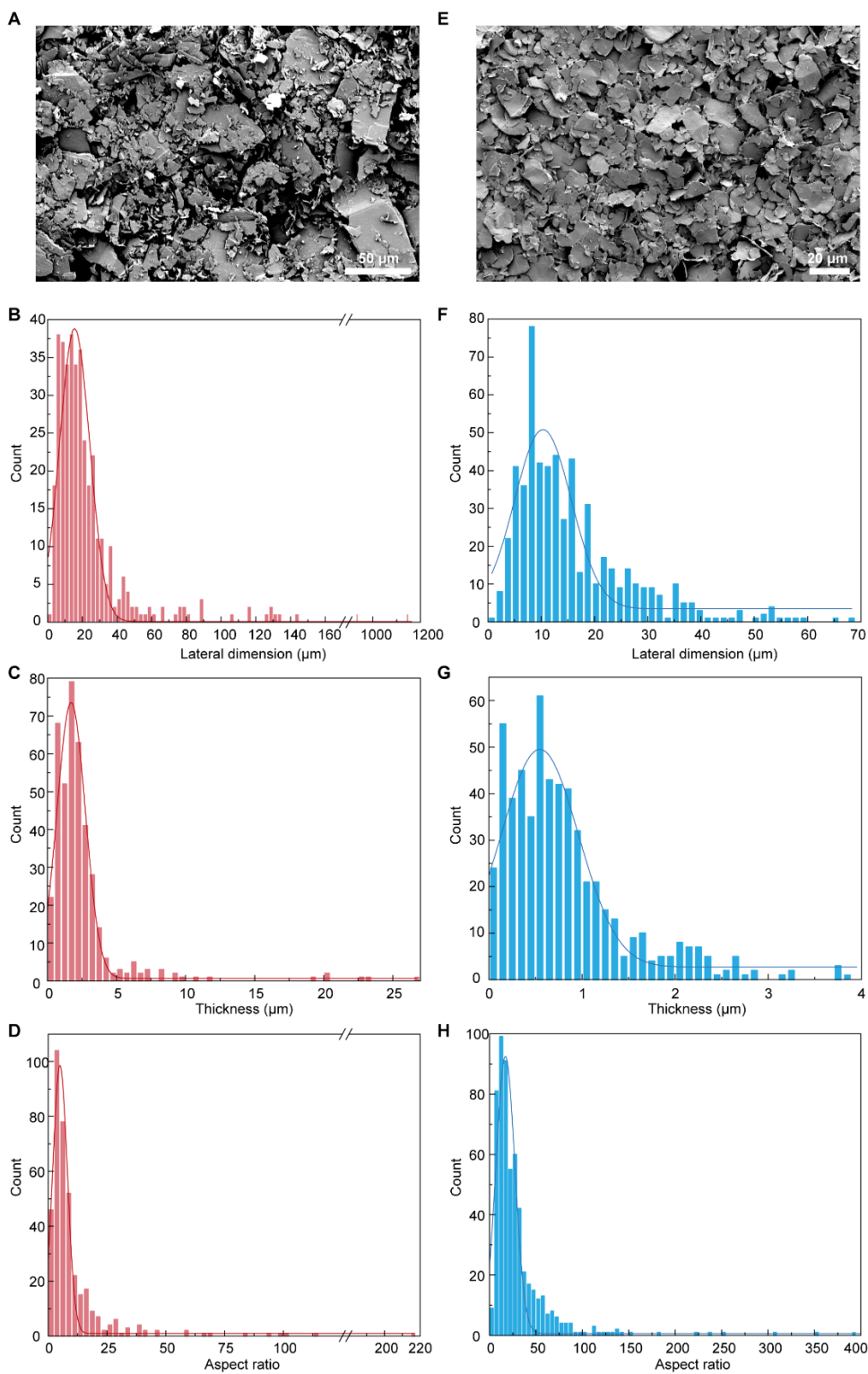

**Figure S1**. Statistical analyses of lateral sizes, thickness, and aspect ratios in "perfect" fillers (graphite) and defective fillers (graphite oxide). (A) SEM images of graphite fillers with a scale bar of 50 μm. (B) Relationships between counts and lateral dimensions of graphite fillers. The data

was plotted with a bin size of 2.5 μm and fitted with Gauss function using Origin Pro software. (C) Relationships between counts and thicknesses of graphite fillers. The data was plotted with a bin size of 0.5 μm and fitted with Gauss function using Origin Pro software. (D) Relationships between counts and aspect ratios of graphite fillers. The aspect ratio of graphite fillers refers to the ratio of lateral dimensions (length and width) to their thickness. The data is plotted with a bin size of 2.5 and fitted with Gauss function using Origin Pro software. (E) SEM images of graphite oxide fillers with a scale bar of 20 μm. (F) Relationships between counts and lateral dimensions of graphite oxide fillers. The data is plotted with a bin size of 1.5 μm and fitted with Gauss function using Origin Pro software. (G) Relationships between counts and thicknesses of graphite fillers. The data is plotted with a bin size of 0.1 μm and fitted with Gauss function using Origin Pro software. (H) Relationships between counts and aspect ratios of graphite oxide fillers. The aspect ratio of graphite fillers refers to the ratio of their lateral dimensions (length and width) to their thickness. The data is plotted with a bin size of 2.5 and fitted with Gauss function using Origin Pro software.

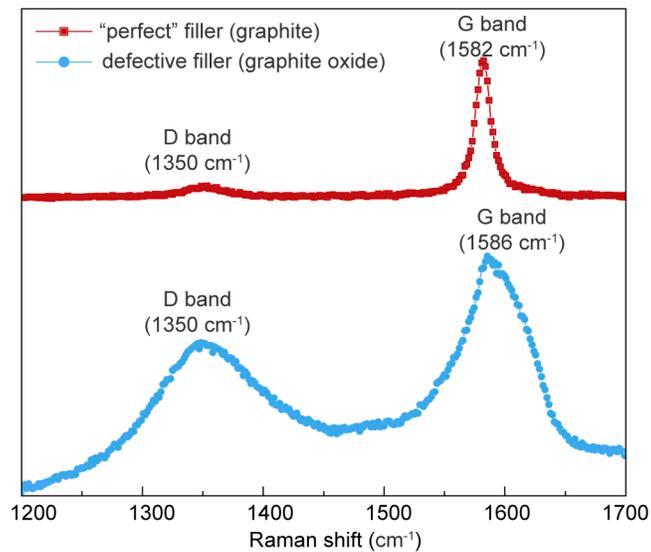

**Figure S2**. Raman spectra of the "perfect" fillers (graphite) and the defective fillers (graphite oxide). The D-band is referred to as the defect band or disordered band.[4-7]

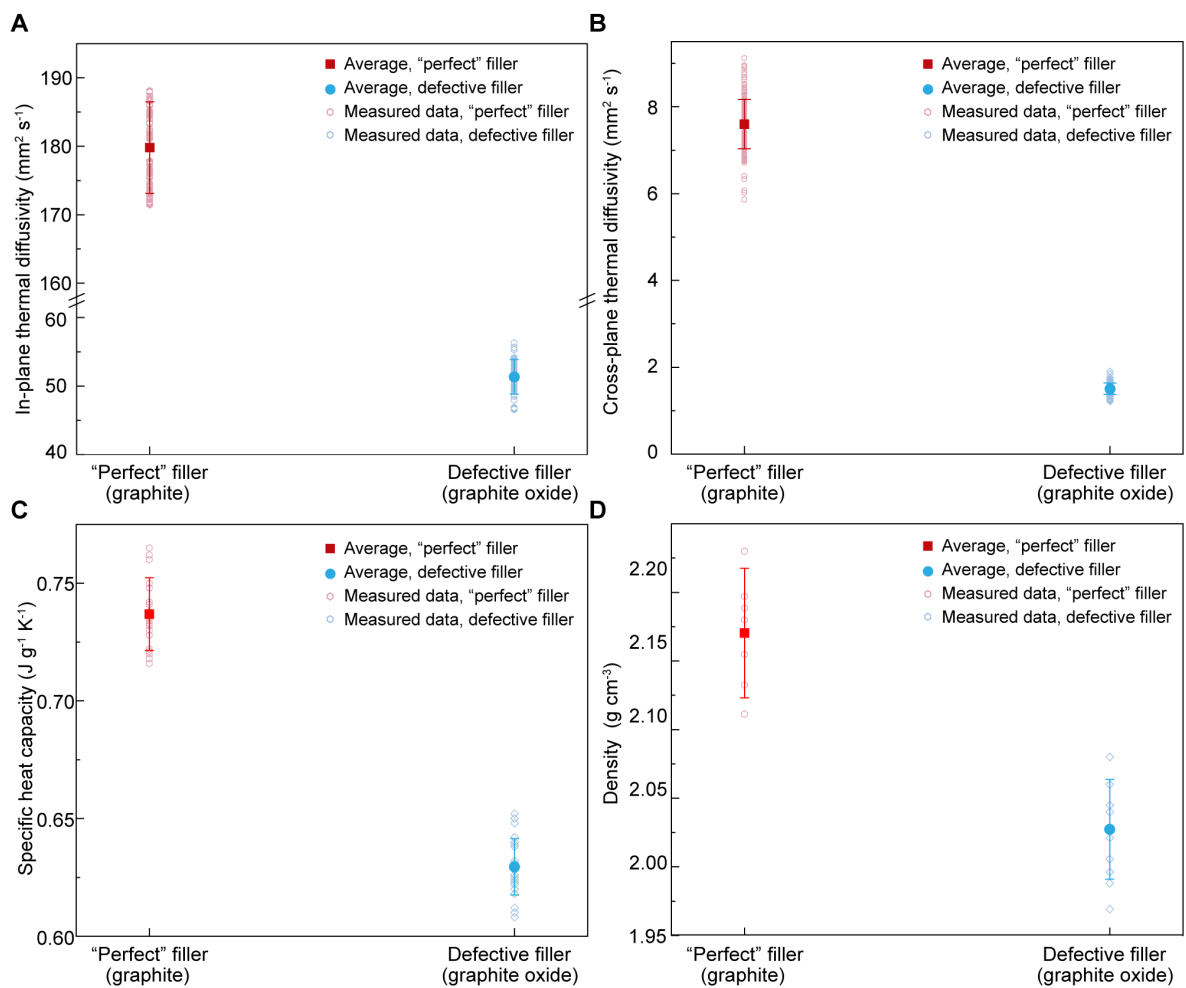

**Figure S3**. (A) Measured in-plane thermal diffusivities of "perfect" fillers (graphite) and defective fillers (graphite oxide) at 25 °C. For details of sample preparations and thermal diffusivity measurements, please refer to the experimental section in supporting information. (B) Measured cross-plane thermal diffusivities of graphite and graphite oxide at 25 °C. For details of sample preparation and thermal diffusivity measurements, please refer to the experimental section in the supporting information. (C) Measured specific heat capacities of compressed graphite and graphite oxide at 25 °C. For details of DSC measurements, please refer to the experimental section in the supporting information. (D) Measured densities of compressed graphite and graphite oxide at 25 °C. The densities are calculated by measuring the thicknesses and weights of the compressed fillers. The error bars in the figure are the population standard deviation based on equations S1 and S2.

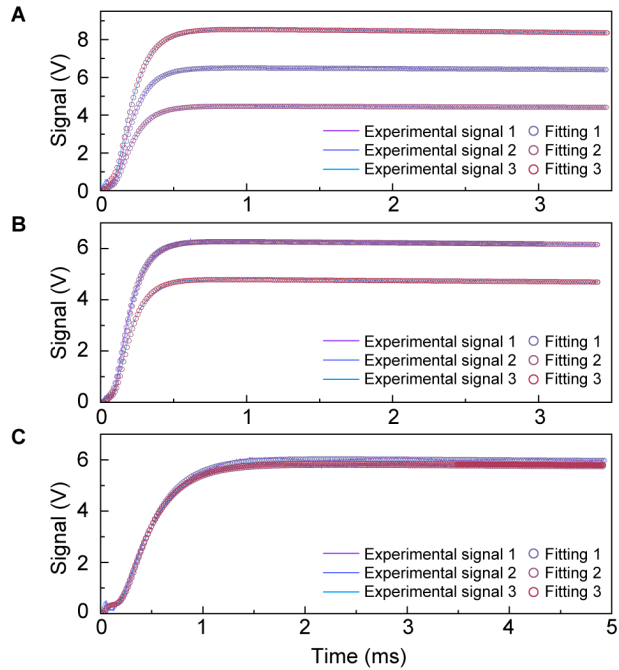

**Figure S4.** Cross-plane thermal diffusivities of PVA films. To minimize random errors and ensure the reproducibility of thermal diffusivity measurements using the laser flash method, we tested the thermal diffusivities of each sample three times. We also measured thermal diffusivities of three different PVA film samples, which we labeled as sample No.1, No.2, and No.3. The "transparent model" in the LFA 467 software was used to fit the experimental signals obtained from the laser flash method. (A) The experimental and fitting results in thermal diffusivity of PVA film (No.1) with a thickness of 0.014 mm. (B) The experimental and fitting results in thermal diffusivity of PVA film (No.2) with a thickness of 0.013 mm. (C) The experimental and fitting results in thermal diffusivity of PVA film (No.3) with a thickness of 0.022 mm.

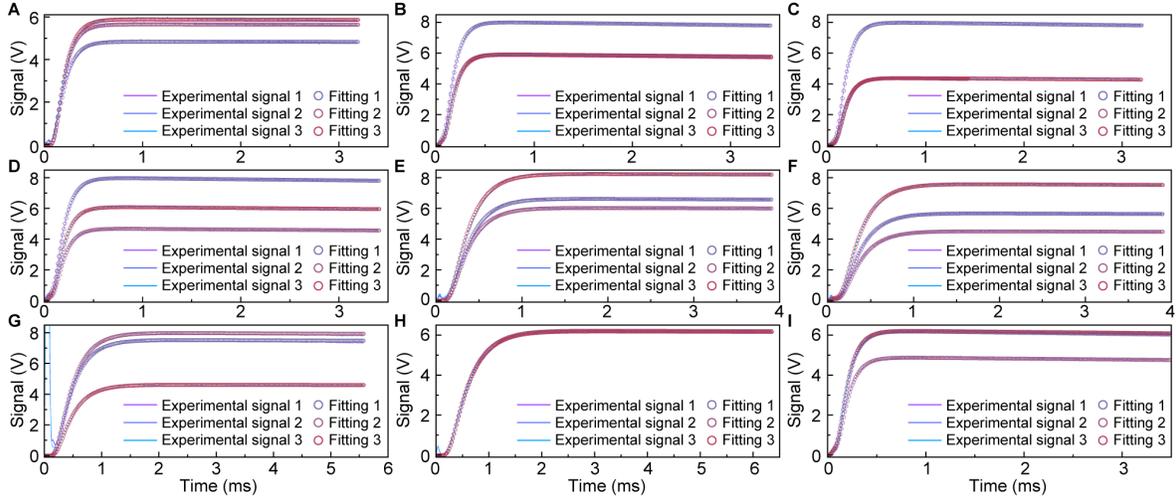

**Figure S5.** Cross-plane thermal diffusivities of PVA/defective filler (1 vol%) composite films. To minimize random errors and ensure the reproducibility of thermal diffusivity measurements using the laser flash method, we tested the thermal diffusivities of each sample three times. We also measured thermal diffusivities of nine different PVA/defective filler (1 vol%) composite samples, which we labeled as sample No.1, No.2, No.3, No.4, No.5, No.6, No.7, No.8, and No.9. The "transparent model" in the LFA 467 software was used to fit the experimental signals obtained from the laser flash method. (A) The experimental and fitting results in thermal diffusivities of sample No.1 with a thickness of 0.015 mm. (B) The experimental and fitting results in thermal diffusivities of sample No.2 with a thickness of 0.016 mm. (C) The experimental and fitting results in thermal diffusivity of sample No.3 with a thickness of 0.015 mm. (D) The experimental and fitting results in thermal diffusivities of sample No.4 with a thickness of 0.017 mm. (E) The experimental and fitting results in thermal diffusivities of sample No.5 with a thickness of 0.014 mm. (F) The experimental and fitting results in thermal diffusivities of sample No.6 with a thickness of 0.014 mm. (G) The experimental and fitting results in thermal diffusivities of sample No.7 with a thickness of 0.015 mm. (H) The experimental and fitting results in thermal diffusivities of sample No.8 with a thickness of 0.015 mm. (I) The experimental and fitting results in thermal diffusivities of sample No.9 with a thickness of 0.017 mm.

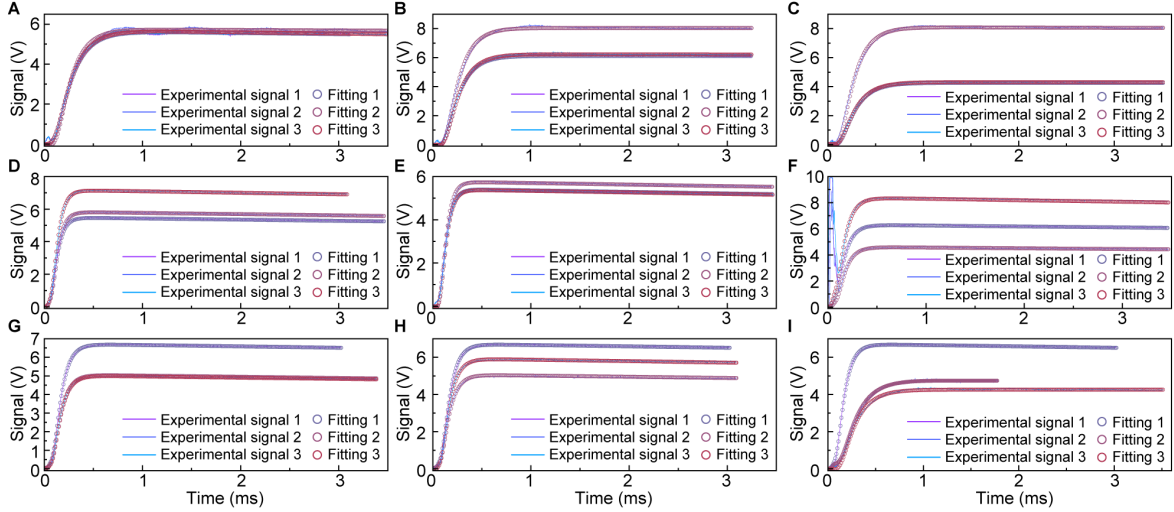

**Figure S6.** Cross-plane thermal diffusivities of PVA/defective filler (3 vol%) composite films. To minimize random errors and ensure the reproducibility of thermal diffusivity measurements using the laser flash method, we tested the thermal diffusivities of each sample three times. We also measured thermal diffusivities of nine different PVA/defective filler (3 vol%) composite samples, which we labeled as sample No.1, No.2, No.3, No.4, No.5, No.6, No.7, No.8, and No.9. The "transparent model" in the LFA 467 software was used to fit the experimental signals obtained from the laser flash method. (A) The experimental and fitting results in thermal diffusivities of sample No.1 with a thickness of 0.025 mm. (B) The experimental and fitting results in thermal diffusivities of sample No.2 with a thickness of 0.024 mm. (C) The experimental and fitting results in thermal diffusivities of sample No.3 with a thickness of 0.025 mm. (D) The experimental and fitting results in thermal diffusivities of sample No.4 with a thickness of 0.016 mm. (E) The experimental and fitting results in thermal diffusivities of sample No.5 with a thickness of 0.016 mm. (F) The experimental and fitting results in thermal diffusivities of sample No.6 with a thickness of 0.018 mm. (G) The experimental and fitting results in thermal diffusivities of sample No.7 with a thickness of 0.021 mm. (H) The experimental and fitting results in thermal diffusivities of sample No.8 with a thickness of 0.017 mm. (I) The experimental and fitting results in thermal diffusivities of sample No.9 with a thickness of 0.015 mm.

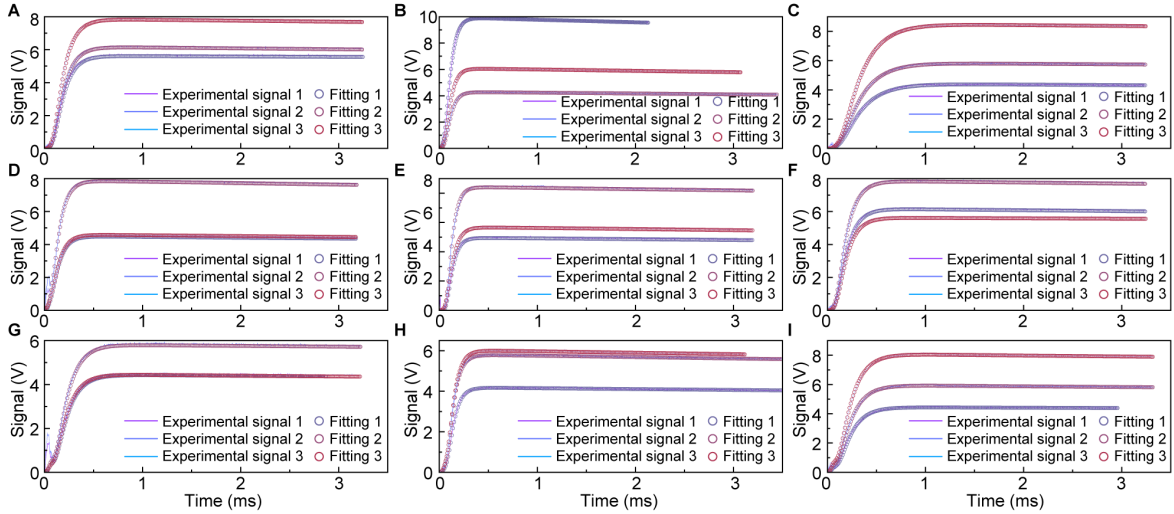

**Figure S7.** Cross-plane thermal diffusivities of PVA/defective filler (5 vol%) composite films. To minimize random errors and ensure the reproducibility of thermal diffusivity measurements using the laser flash method, we tested the thermal diffusivities of each sample three times. We also measured thermal diffusivities of nine different PVA/defective filler (5 vol%) composite samples, which we labeled as sample No.1, No.2, No.3, No.4, No.5, No.6, No.7, No.8, and No.9. The "transparent model" in the LFA 467 software was used to fit the experimental signals obtained from the laser flash method. (A) The experimental and fitting results in thermal diffusivities of sample No.1 with a thickness of 0.030 mm. (B) The experimental and fitting results in thermal diffusivities of sample No.2 with a thickness of 0.019 mm. (C) The experimental and fitting results in thermal diffusivities of sample No.3 with a thickness of 0.031 mm. (D) The experimental and fitting results in thermal diffusivities of sample No.4 with a thickness of 0.028 mm. (E) The experimental and fitting results in thermal diffusivities of sample No.5 with thickness of 0.021 mm. (F) The experimental and fitting results in thermal diffusivities of sample No.6 with thickness of 0.029 mm. (G) The experimental and fitting results in thermal diffusivities of sample No.7 with a thickness of 0.031 mm. (H) The experimental and fitting results in thermal diffusivities of sample No.8 with a thickness of 0.023 mm. (I) The experimental and fitting results in thermal diffusivities of sample No.9 with a thickness of 0.032 mm.

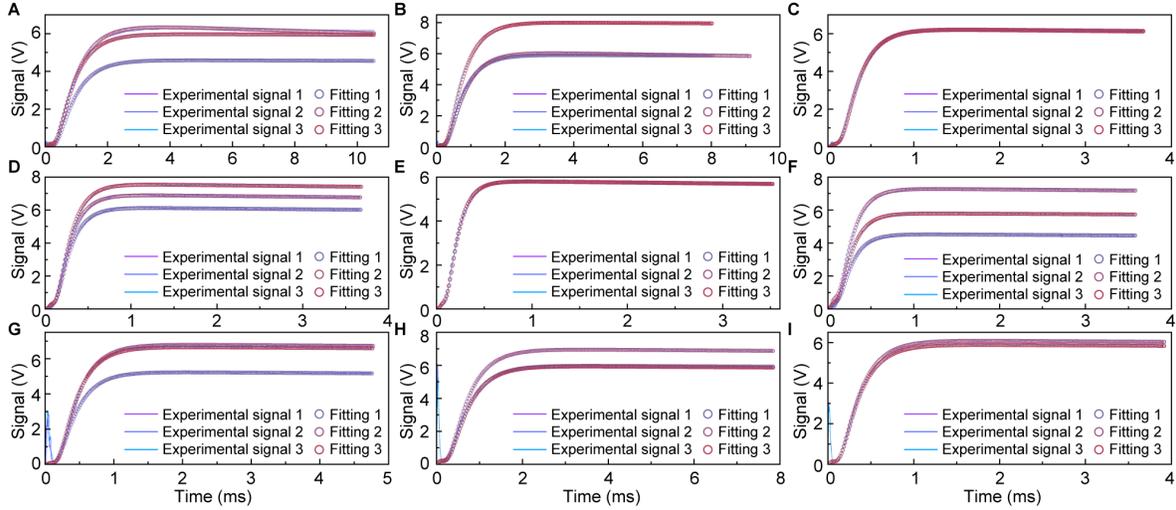

**Figure S8.** Cross-plane thermal diffusivities of PVA/"perfect" filler (1 vol%) composite films. To minimize random errors and ensure the reproducibility of thermal diffusivity measurements using the laser flash method, we tested the thermal diffusivities of each sample three times. We also measured thermal diffusivities of nine different PVA/"perfect" filler (1 vol%) composite samples, which we labeled as sample No.1, No.2, No.3, No.4, No.5, No.6, No.7, No.8, and No.9. The "transparent model" in the LFA 467 software was used to fit the experimental signals obtained from the laser flash method. (A) The experimental and fitting results in thermal diffusivity of polymer composite film (No.1) with thickness of 0.019 mm. (B) The experimental and fitting results in thermal diffusivity of polymer composite film (No.2) with thickness of 0.018 mm. (C) The experimental and fitting results in thermal diffusivity of polymer composite film (No.3) with thickness of 0.016 mm. (D) The experimental and fitting results in thermal diffusivity of polymer composite film (No.4) with thickness of 0.034 mm. (E) The experimental and fitting results in thermal diffusivity of polymer composite film (No.5) with thickness of 0.017 mm. (F) The experimental and fitting results in thermal diffusivity of polymer composite film (No.6) with thickness of 0.029 mm. (G) The experimental and fitting results in thermal diffusivity of polymer composite film (No.7) with thickness of 0.022 mm. (H) The experimental and fitting results in thermal diffusivity of polymer composite film (No.8) with thickness of 0.028 mm. (I) The experimental and fitting results in thermal diffusivity of polymer composite film (No.9) with thickness of 0.030 mm.

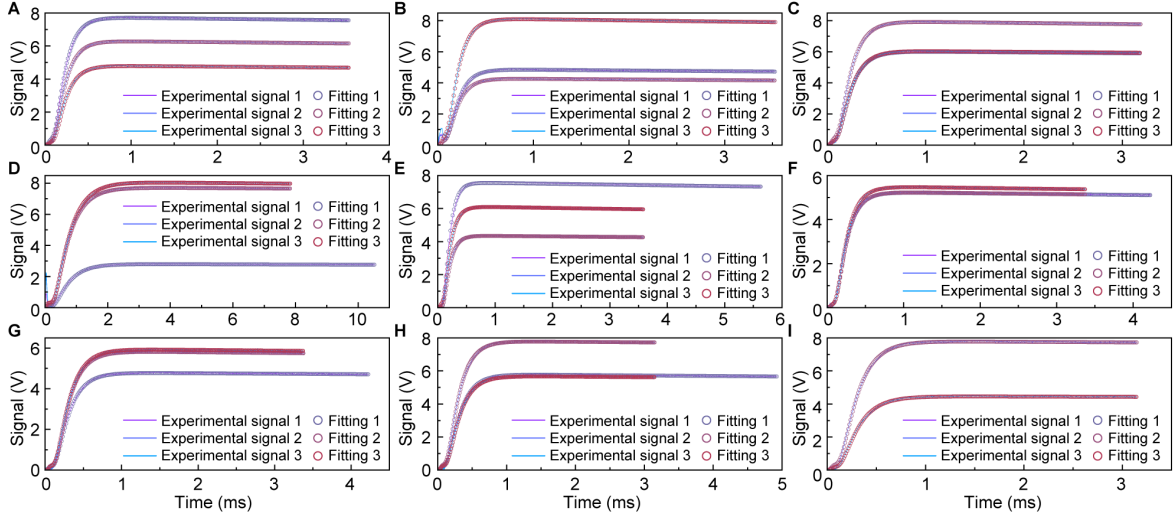

**Figure S9.** Cross-plane thermal diffusivities of PVA/"perfect" filler (3 vol%) composite films. To minimize random errors and ensure the reproducibility of thermal diffusivity measurements using the laser flash method, we tested the thermal diffusivities of each sample three times. We also measured thermal diffusivities of nine different PVA/"perfect" filler (3 vol%) composite samples, which we labeled as sample No.1, No.2, No.3, No.4, No.5, No.6, No.7, No.8, and No.9. The "transparent model" in the LFA 467 software was used to fit the experimental signals obtained from the laser flash method. (A) The experimental and fitting results in thermal diffusivity of polymer composite film (No.1) with thickness of 0.020 mm. (B) The experimental and fitting results in thermal diffusivity of polymer composite film (No.2) with thickness of 0.020 mm. (C) The experimental and fitting results in thermal diffusivity of polymer composite film (No.3) with thickness of 0.022 mm. (D) The experimental and fitting results in thermal diffusivity of polymer composite film (No.4) with thickness of 0.036 mm. (E) The experimental and fitting results in thermal diffusivity of polymer composite film (No.5) with thickness of 0.018 mm. (F) The experimental and fitting results in thermal diffusivity of polymer composite film (No.6) with thickness of 0.020 mm. (G) The experimental and fitting results in thermal diffusivity of polymer composite film (No.7) with thickness of 0.022 mm. (H) The experimental and fitting results in thermal diffusivity of polymer composite film (No.8) with thickness of 0.023 mm. (I) The experimental and fitting results in thermal diffusivity of polymer composite film (No.9) with thickness of 0.023 mm.

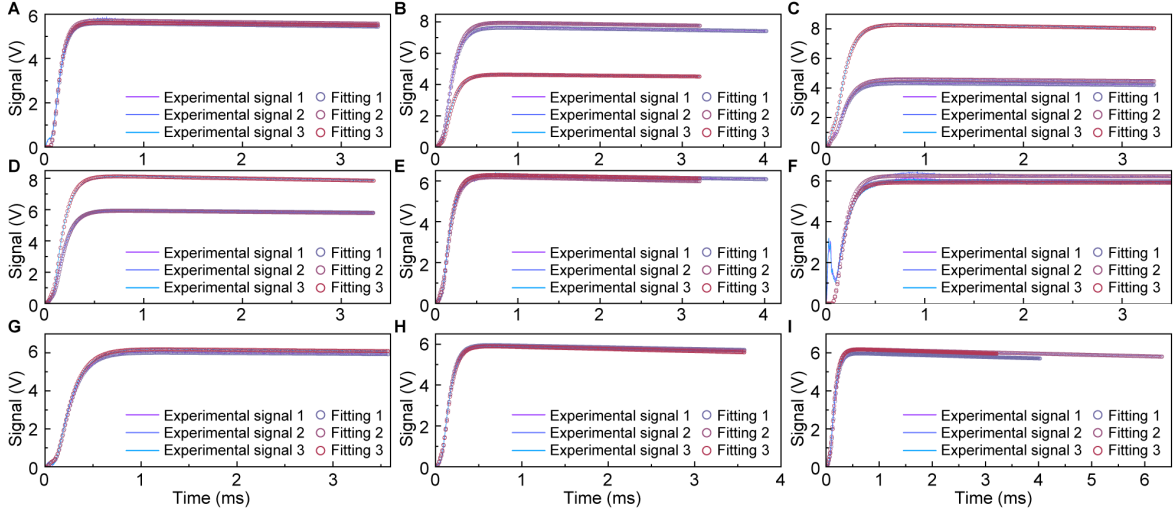

**Figure S10.** Cross-plane thermal diffusivities of PVA/"perfect" filler (5 vol%) composite films. To minimize random errors and ensure the reproducibility of thermal diffusivity measurements using the laser flash method, we tested the thermal diffusivities of each sample three times. We also measured thermal diffusivities of nine different PVA/"perfect" filler (5 vol%) composite samples, which we labeled as sample No.1, No.2, No.3, No.4, No.5, No.6, No.7, No.8, and No.9. The "transparent model" in the LFA 467 software was used to fit the experimental signals obtained from the laser flash method. (A) The experimental and fitting results in thermal diffusivity of polymer composite film (No.1) with thickness of 0.019 mm. (B) The experimental and fitting results in thermal diffusivity of polymer composite film (No.2) with thickness of 0.021 mm. (C) The experimental and fitting results in thermal diffusivity of polymer composite film (No.3) with thickness of 0.022 mm. (D) The experimental and fitting results in thermal diffusivity of polymer composite film (No.4) with thickness of 0.022 mm. (E) The experimental and fitting results in thermal diffusivity of polymer composite film (No.5) with thickness of 0.021 mm. (F) The experimental and fitting results in thermal diffusivity of polymer composite film (No.6) with thickness of 0.022 mm. (G) The experimental and fitting results in thermal diffusivity of polymer composite film (No.7) with thickness of 0.026 mm. (H) The experimental and fitting results in thermal diffusivity of polymer composite film (No.8) with thickness of 0.020 mm. (I) The experimental and fitting results in thermal diffusivity of polymer composite film (No.9) with thickness of 0.022 mm.

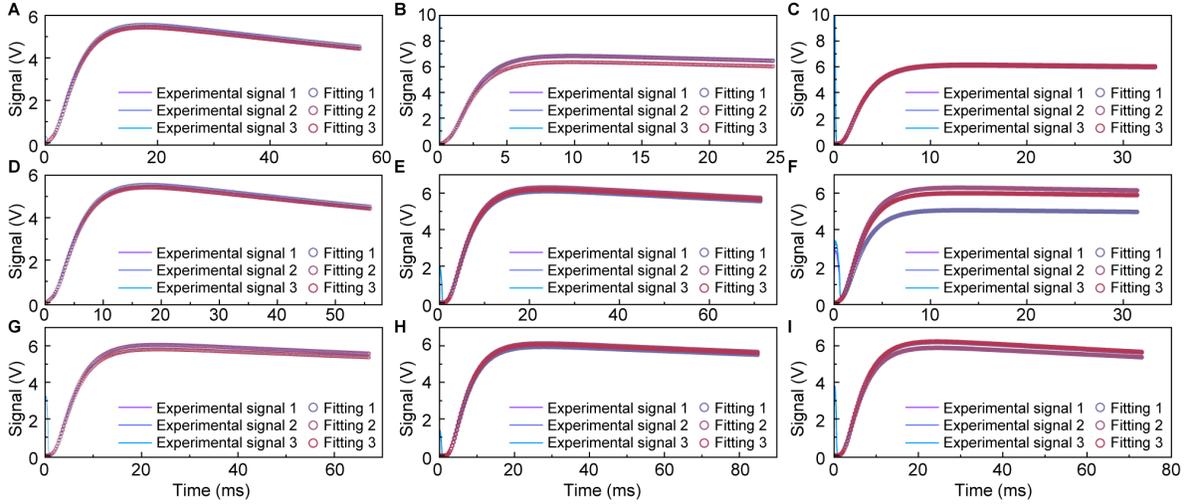

**Figure S11.** Cross-plane thermal diffusivities of compressed graphite oxide. To minimize random errors and ensure the reproducibility of thermal diffusivity measurements using the laser flash method, we tested the thermal diffusivities of each sample three times. We also measured thermal diffusivities of nine different compressed graphite oxide samples, which we labeled as sample No.1, No.2, No.3, No.4, No.5, No.6, No.7, No.8, and No.9. The "penetration model" in the LFA 467 software was used to fit the cross-plane thermal diffusivity experimental signals obtained from the laser flash method. (A) The experimental and fitting results in cross-plane thermal diffusivities of compressed graphite oxide (No.1) with thickness of 0.233 mm. (B) The experimental and fitting results in cross-plane thermal diffusivities of compressed graphite oxide (No.2) with thickness of 0.138 mm. (C) The experimental and fitting results in cross-plane thermal diffusivities of compressed graphite oxide (No.3) with thickness of 0.181 mm. (D) The experimental and fitting results in cross-plane thermal diffusivities of compressed graphite oxide (No.4) with thickness of 0.250 mm. (E) The experimental and fitting results in cross-plane thermal diffusivities of compressed graphite oxide (No.5) with thickness of 0.260 mm. (F) The experimental and fitting results in cross-plane thermal diffusivities of compressed graphite oxide (No.6) with thickness of 0.171 mm. (G) The experimental and fitting results in cross-plane thermal diffusivities of compressed graphite oxide (No.7) with thickness of 0.254 mm. (H) The experimental and fitting results in cross-plane thermal diffusivities of compressed graphite oxide (No.8) with thickness of 0.281 mm. (I) The experimental and fitting results in cross-plane thermal diffusivities of compressed graphite oxide (No.9) with thickness of 0.262 mm. For details of sample preparations and thermal diffusivity measurements, please refer to the experimental section in the supporting information.

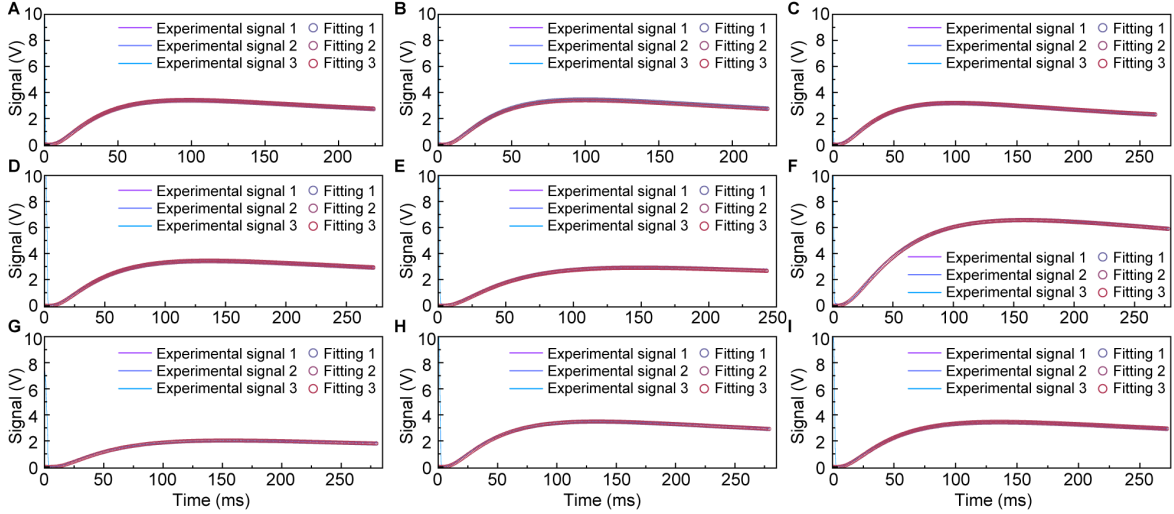

**Figure S12.** In-plane thermal diffusivities of compressed graphite oxide. To minimize random errors and ensure the reproducibility of thermal diffusivity measurements using the laser flash method, we tested the thermal diffusivities of each sample three times. We also measured thermal diffusivities of nine different compressed graphite oxide samples, which we labeled as sample No.1, No.2, No.3, No.4, No.5, No.6, No.7, No.8, and No.9. The "in-plane anisotropic model" in the LFA 467 software was used to fit the in-plane thermal diffusivity experimental signals obtained from the laser flash method. (A) The experimental and fitting results in in-plane thermal diffusivities of compressed graphite oxide (No.1) with thickness of 0.195 mm. (B) The experimental and fitting results in in-plane thermal diffusivities of compressed graphite oxide (No.2) with thickness of 0.153 mm. (C) The experimental and fitting results in in-plane thermal diffusivities of compressed graphite oxide (No.3) with thickness of 0.236 mm. (D) The experimental and fitting results in in-plane thermal diffusivities of compressed graphite oxide (No.4) with thickness of 0.263 mm. (E) The experimental and fitting results in in-plane thermal diffusivities of compressed graphite oxide (No.5) with thickness of 0.203 mm. (F) The experimental and fitting results in in-plane thermal diffusivities of compressed graphite oxide (No.6) with thickness of 0.212 mm. (G) The experimental and fitting results in in-plane thermal diffusivities of compressed graphite oxide (No.7) with thickness of 0.214 mm. (H) The experimental and fitting results in in-plane thermal diffusivities of compressed graphite oxide (No.8) with thickness of 0.230 mm. (I) The experimental and fitting results in in-plane thermal diffusivities of compressed graphite oxide (No.9) with thickness of 0.214 mm. For details of sample preparations and thermal diffusivity measurements, please refer to the experimental section in the supporting information.

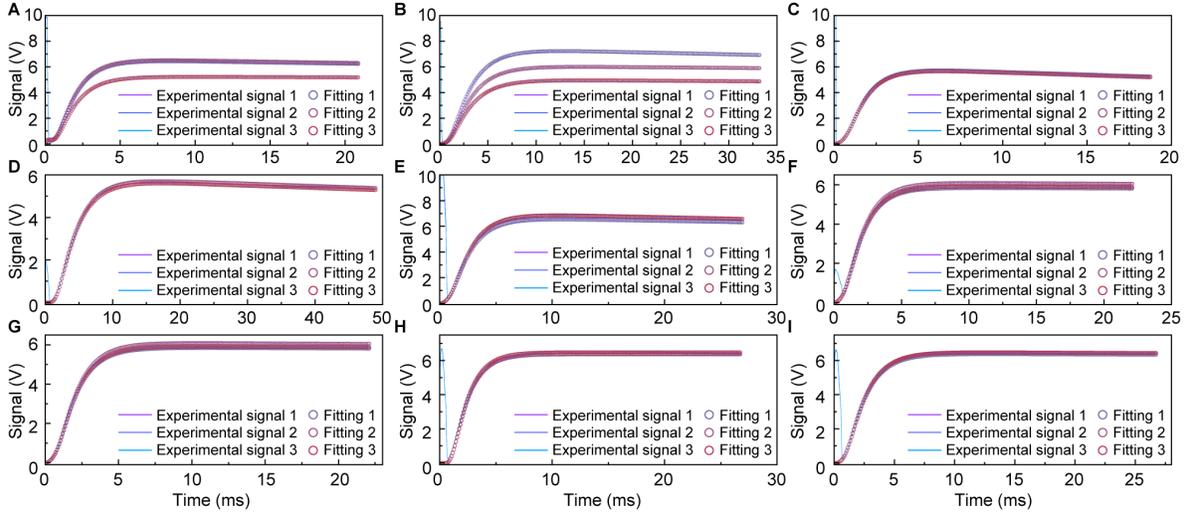

**Figure S13.** Cross-plane thermal diffusivities of compressed graphite. To minimize random errors and ensure the reproducibility of thermal diffusivity measurements using the laser flash method, we tested the thermal diffusivities of each sample three times. We also measured thermal diffusivities of nine different compressed graphite samples, which we labeled as sample No.1, No.2, No.3, No.4, No.5, No.6, No.7, No.8, and No.9. The "penetration model" in the LFA 467 software was used to fit the cross-plane thermal diffusivity experimental signals obtained from the laser flash method. (A) The experimental and fitting results in cross-plane thermal diffusivities of compressed graphite (No.1) with thickness of 0.312 mm. (B) The experimental and fitting results in cross-plane thermal diffusivities of compressed graphite (No.2) with thickness of 0.399 mm. (C) The experimental and fitting results in cross-plane thermal diffusivities of compressed graphite (No.3) with thickness of 0.296 mm. (D) The experimental and fitting results in cross-plane thermal diffusivities of compressed graphite (No.4) with thickness of 0.462 mm. (E) The experimental and fitting results in cross-plane thermal diffusivities of compressed graphite (No.5) with thickness of 0.362 mm. (F) The experimental and fitting results in cross-plane thermal diffusivities of compressed graphite (No.6) with thickness of 0.316 mm. (G) The experimental and fitting results in cross-plane thermal diffusivities of compressed graphite (No.7) with thickness of 0.327 mm. (H) The experimental and fitting results in cross-plane thermal diffusivities of compressed graphite (No.8) with thickness of 0.335 mm. (I) The experimental and fitting results in cross-plane thermal diffusivities of compressed graphite (No.9) with thickness of 0.345 mm. For details of sample preparations and thermal diffusivity measurements, please refer to the experimental section in the supporting information.

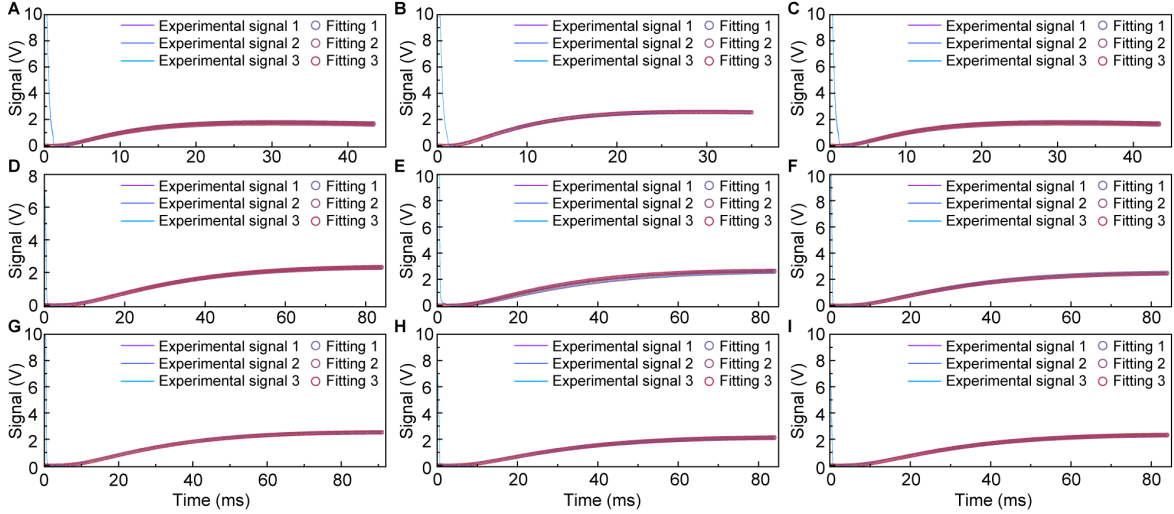

**Figure S14.** In-plane thermal diffusivities of compressed graphite. To minimize random errors and ensure the reproducibility of thermal diffusivity measurements using the laser flash method, we tested the thermal diffusivities of each sample three times. We also measured thermal diffusivities of nine different compressed graphite samples, which we labeled as sample No.1, No.2, No.3, No.4, No.5, No.6, No.7, No.8, and No.9. The "in-plane anisotropic model" was used to fit the in-plane thermal diffusivity experimental signals obtained from the laser flash method. (A) The experimental and fitting results in in-plane thermal diffusivities of compressed graphite (No.1) with thickness of 0.463 mm. (B) The experimental and fitting results in in-plane thermal diffusivities of compressed graphite (No.2) with thickness of 0.296 mm. (C) The experimental and fitting results in in-plane thermal diffusivities of compressed graphite (No.3) with thickness of 0.381 mm. (D) The experimental and fitting results in in-plane thermal diffusivities of compressed graphite (No.4) with thickness of 0.277 mm. (E) The experimental and fitting results in in-plane thermal diffusivities of compressed graphite (No.5) with thickness of 0.254 mm. (F) The experimental and fitting results in in-plane thermal diffusivities of compressed graphite (No.6) with thickness of 0.226 mm. (G) The experimental and fitting results in in-plane thermal diffusivities of compressed graphite (No.7) with thickness of 0.261 mm. (H) The experimental and fitting results in in-plane thermal diffusivities of compressed graphite (No.8) with thickness of 0.279 mm. (I) The experimental and fitting results in in-plane thermal diffusivities of compressed graphite (No.9) with thickness of 0.246 mm. For details of sample preparations and thermal diffusivity measurements, please refer to the experimental section in the supporting information.

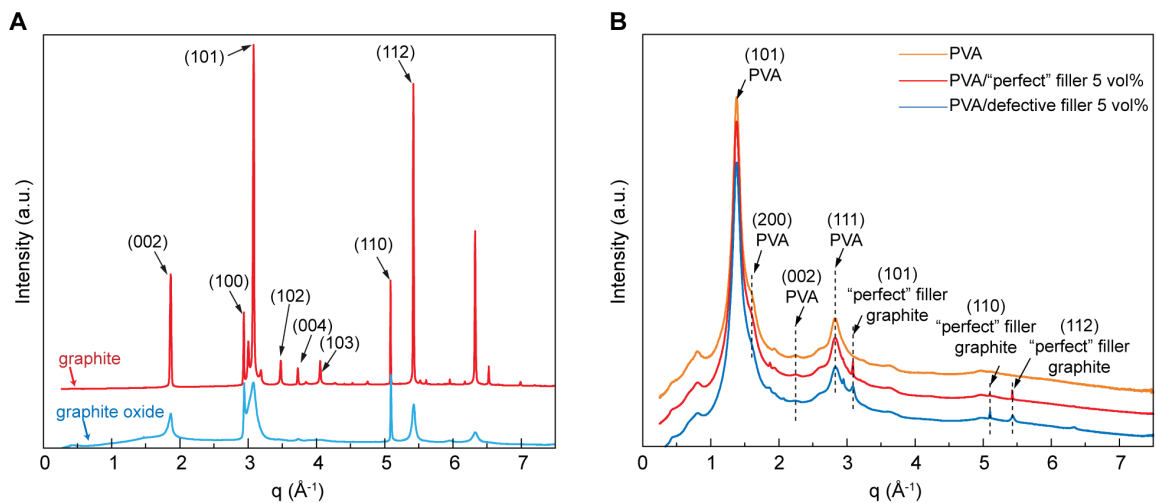

**Figure S15.** The one-dimensional 1D curves of the synchrotron X-ray scattering intensity versus scattering vector (q) for several materials: compressed graphite, compressed graphite oxide, PVA film. PVA/"perfect" filler (5 vol %) composite films, and PVA/defective filler (5 vol %) composite films. (A) Compressed commercial graphite (CAS: 7782-42-5) purchased from Sigma-Aldrich and compressed graphite oxide made by Hummer's method.[1-3] (B) PVA films, PVA/"perfect" filler (5 vol %) composite films, and PVA/defective filler (5 vol %) composite film.[8, 9] The preparation details for making samples for synchrotron X-ray scattering measurements are in the Section 1 in the supporting information.

## Section 3. Effective theory of dynamical defects enhanced lattice thermal conductivity.

### 3.1 Effective quantum mechanical model (vibrational Hamiltonian model) setup.

We consider a phonon bath $\omega_q$ as the backbone vibrational modes in the polymer, which interacts with the defects. Rather than treating the defect as a static object, here we treat the defect as a two-level system, resembling PW Anderson's treatment on the specific heat on amorphous solids,[10] with the two levels from the vibrational states of the local defects as $\omega_1$ and $\omega_2$. The total Hamiltonian can be written as

$$H = \sum_{\mathbf{q}} \omega_{\mathbf{q}} \left(a_{\mathbf{q}}^+ a_{\mathbf{q}} + \frac{1}{2}\right) + \omega_1 \left(a_1^+ a_1 + \frac{1}{2}\right) + \omega_2 \left(a_2^+ a_2 + \frac{1}{2}\right)$$
$$+ \sum_{\mathbf{q}j}(V_{\mathbf{q}j} a_{\mathbf{q}}^+ a_j + V_{\mathbf{q}j}^* a_j^+ a_{\mathbf{q}}) + V_{12} a_1^+ a_2 + V_{12}^* a_2^+ a_1 \text{ (Equation S3.1.1)}$$

Where the coefficients $V_{\mathbf{q}j}$ denotes the hybridization element between the vibrational defects level $j=1,2$ to the polymer $\mathbf{q}$, and $V_{12}$ denotes the tunneling between the two defect modes. In Anderson's work,[10] from a tunneling picture, it is shown that $V_{12} = \omega_1 \, exp\left(-\sqrt{\frac{1}{2}mV\Delta x}\right)$.[10]

Here, all operators satisfy the Bosonic canonical commutation relation as

$$[a_m, a_n^+] = \delta_{mn}, m, n = \mathbf{q}, 1, 2 \text{ (Equation S3.1.2)}$$

### 3.2 Green's functions.

Since Equation S3.1.1 is quadratic, we anticipate the Green's functions have a closed form solution. We define the general retarded Green's functions for both the polymer and the defects as

$$D_{mn}(t - t') = -i\theta(t - t')\langle[a_m(t), a_n^+(t')]\rangle \text{ (Equation S3.2.1)}$$

Where $m, n = \mathbf{q}, 1, 2$. We adopt the equation of motion method to obtain the Green's functions. Taking time derivative to Equation S3.2.1, we have

$$i\partial_t D_{mn}(t - t') = \delta(t - t')\delta_{mn} + i\theta(t - t')\langle[[H, a_m](t), a_n^+(t')]\rangle \text{ (Equation S3.2.2)}$$

Where the Heisenberg equation-of-motion is used. Now, using the full Hamiltonian Equation S3.1.1, we compute the commutator $[H, a_m]$; for any $m \in \{\{\mathbf{q}\}, 1, 2\}$, complement $\tilde{m} \equiv \{\{\mathbf{q}\}, 1, 2\} - m$

$$[H, a_m] = -\omega_m a_m - \sum_{\tilde{m}} V_{m\tilde{m}} a_{\tilde{m}} \text{ (Equation S3.2.3)}$$

Substituting Equation S3.2.3 back to Equation S3.2.2, we have the general equation-of-motion of Green's functions:

$$i\partial_t D_{mn}(t-t') = \delta(t-t')\delta_{mn} + \omega_m D_{mn}(t-t') + \sum_{\tilde{m}} V_{m\tilde{m}} D_{\tilde{m}n}(t-t') \quad \text{(Equation S3.2.4)}$$

Defining the frequency-domain Green's functions as

$$D_{mn}(\omega) = \int d(t-t') e^{+i(\omega+i0^+)(t-t')} D_{mn}(t-t') \quad \text{(Equation S3.2.5)}$$

Then the Green's function equation of motion Equation S3.2.4 can be rewritten as

$$(\omega - \omega_m + i0^+) D_{mn}(\omega) - \sum_{\tilde{m}} V_{m\tilde{m}} D_{\tilde{m}n}(\omega) = \delta_{mn} \quad \text{(Equation S3.2.6)}$$

Or equivalently, writing in explicit form for $m \in \{\{q\}, 1, 2\}$, we have

$$(\omega - \omega_q + i0^+) D_{qk}(\omega) - V_{q1} D_{1k}(\omega) - V_{q2} D_{2k}(\omega) = \delta_{kq}$$
$$(\omega - \omega_q + i0^+) D_{q1}(\omega) - V_{q1} D_{11}(\omega) - V_{q2} D_{21}(\omega) = 0$$
$$(\omega - \omega_q + i0^+) D_{q2}(\omega) - V_{q2} D_{22}(\omega) - V_{q1} D_{12}(\omega) = 0$$
$$(\omega - \omega_1 + i0^+) D_{1q}(\omega) - \sum_k V_{1k} D_{kq}(\omega) - V_{12} D_{2q}(\omega) = 0$$
$$(\omega - \omega_1 + i0^+) D_{11}(\omega) - \sum_k V_{1k} D_{k1}(\omega) - V_{12} D_{21}(\omega) = 1 \quad \text{(Equation S3.2.7)}$$
$$(\omega - \omega_1 + i0^+) D_{12}(\omega) - \sum_k V_{1k} D_{k2}(\omega) - V_{12} D_{22}(\omega) = 0$$
$$(\omega - \omega_2 + i0^+) D_{2q}(\omega) - \sum_k V_{2k} D_{kq}(\omega) - V_{21} D_{1q}(\omega) = 0$$
$$(\omega - \omega_2 + i0^+) D_{21}(\omega) - \sum_k V_{2k} D_{k1}(\omega) - V_{21} D_{11}(\omega) = 0$$
$$(\omega - \omega_2 + i0^+) D_{22}(\omega) - \sum_k V_{2k} D_{k2}(\omega) - V_{21} D_{12}(\omega) = 1$$

This is a set of 9 linear equations with 9 unknown variables, and thus can be solved out explicitly. For later computational convenience, we also need a set of adjoint equations. This can be obtained by either taking the derivative w.r.t. $t'$ in Equation S3.2.1, or using the Lehmann representation of Green's function, which is more general. The frequency-domain Green's functions can be written as

$$D_{mn}(\omega) = \frac{1}{Z} \sum_{pq} \left[ \frac{e^{-\beta E_p} \langle p|a_m|q\rangle\langle q|a_n^+|p\rangle}{\omega + E_p - E_q + i0^+} - \frac{e^{-\beta E_p} \langle p|a_n^+|q\rangle\langle q|a_m|p\rangle}{\omega + E_q - E_p + i0^+} \right] \quad \text{(Equation S3.2.8)}$$

from which we immediately obtain $D_{mn}(\omega) = D_{nm}^*(\omega)$. Substituting into Equation S3.2.6, we obtain

$$(\omega - \omega_n - i0^+) D_{mn}(\omega) - \sum_{\tilde{n}} V_{n\tilde{n}}^* D_{m\tilde{n}}(\omega) = \delta_{mn} \quad \text{(Equation S3.2.9)}$$

Explicitly, we have two of the following equations that come handy later:

$$D_{\alpha k}(\omega) = \frac{V_{k1}^* D_{\alpha 1}(\omega) + V_{k2}^* D_{\alpha 2}(\omega)}{\omega - \omega_k - i0^+} = \sum_{\beta=1,2} \frac{V_{k\beta}^* D_{\alpha\beta}(\omega)}{\omega - \omega_k - i0^+}, \alpha = 1,2 \quad \text{(Equation S3.2.10)}$$

### 3.3 Phonon Green's function in polymer matrix.

We now are ready to compute the phonon Green's function in the polymer $D_{qk}(\omega)$. Label above 9 equations in Equation (S3.2.7) as (a)-(i). Equations (b) and (c) can be rewritten as

$$D_{k1}(\omega) = \frac{V_{k1} D_{11}(\omega) + V_{k2} D_{21}(\omega)}{\omega - \omega_k + i0^+}$$

$$D_{k2}(\omega) = \frac{V_{k2} D_{22}(\omega) + V_{k1} D_{12}(\omega)}{\omega - \omega_k + i0^+} \quad \text{(Equation S3.3.1)}$$

Substituting the Equation S3.3.1 back to Equations (e) and (h) to cancel out $D_{k1}(\omega)$, and similarly back to (f) and (i) to cancel out $D_{k2}(\omega)$, and define coefficients

$$\lambda_{11}(\omega) = \sum_k \frac{V_{1k} V_{k1}}{\omega - \omega_k + i0^+}, \lambda_{12}(\omega) = \sum_k \frac{V_{1k} V_{k2}}{\omega - \omega_k + i0^+},$$

$$\lambda_{21}(\omega) = \sum_k \frac{V_{2k} V_{k1}}{\omega - \omega_k + i0^+}, \lambda_{22}(\omega) = \sum_k \frac{V_{2k} V_{k2}}{\omega - \omega_k + i0^+} \quad \text{(Equation S3.3.2)}$$

Then we have the following equations only containing the dynamical defects' Green's functions:

$$(\omega - \omega_1 - \lambda_{11}(\omega))D_{11}(\omega) - (\lambda_{12}(\omega) + V_{12})D_{21}(\omega) = 1$$
$$(\omega - \omega_2 - \lambda_{22}(\omega))D_{21}(\omega) - (\lambda_{21}(\omega) + V_{21})D_{11}(\omega) = 0$$
$$(\omega - \omega_1 - \lambda_{11}(\omega))D_{12}(\omega) - (\lambda_{12}(\omega) + V_{12})D_{22}(\omega) = 0 \quad \text{(Equation S3.3.3)}$$
$$(\omega - \omega_2 - \lambda_{22}(\omega))D_{22}(\omega) - (\lambda_{21}(\omega) + V_{21})D_{12}(\omega) = 1$$

Solving linear equation of Equation S3.3.3, we have defect Green's functions

$$D_{11}(\omega) = -\frac{\omega - \omega_2 - \lambda_{22}(\omega)}{g(\omega)}, D_{12}(\omega) = -\frac{\lambda_{12}(\omega) + V_{12}}{g(\omega)}$$

$$D_{21}(\omega) = -\frac{\lambda_{21}(\omega) + V_{21}}{g(\omega)}, D_{22}(\omega) = -\frac{\omega - \omega_1 - \lambda_{11}(\omega)}{g(\omega)} \quad \text{(Equation S3.3.4)}$$

Where $g(\omega) \equiv (\lambda_{12}(\omega) + V_{12})(\lambda_{21}(\omega) + V_{21}) - (\omega - \omega_1 - \lambda_{11}(\omega))(\omega - \omega_2 - \lambda_{22}(\omega))$.

Now substituting Equation S3.2.10 back to Equation (a) in Equation S3.2.7, we have

$$D_{qk}(\omega) = \frac{\delta_{qk}}{\omega - \omega_q + i0^+} + \frac{\sum_{\alpha,\beta=1}^{2} V_{q\alpha} V_{k\beta}^* D_{\alpha\beta}(\omega)}{(\omega - \omega_q + i0^+)(\omega - \omega_k - i0^+)} \quad \text{(Equation S3.3.5)}$$

Or write down explicitly using Equation S3.3.4, we have the final form

$$D_{qk}(\omega)$$
$$= \frac{\delta_{qk}}{\omega - \omega_q}$$
$$- \frac{1}{(\omega - \omega_q)(\omega - \omega_k)g(\omega)} \begin{pmatrix} V_{q1}V_{k1}^*(\omega - \omega_2 - \lambda_{22}(\omega)) + V_{q1}V_{k2}^*(\lambda_{12}(\omega) + V_{12}) \\ +V_{q2}V_{k1}^*(\lambda_{21}(\omega) + V_{21}) + V_{q2}V_{k2}^*(\omega - \omega_1 - \lambda_{11}(\omega)) \end{pmatrix}$$ (Equation S3.3.6)

which is the Green's function of the polymer phonon dispersion $\omega_q$, after interacting with the dynamical defects with two energy levels.

Consider the case where the two levels of defects hybridize with the polymer backbone are the same, so $V_{k1}^* = V_{k2}^* = V_k^*$, $V_{k1} = V_{k2} = V_k$, $\lambda_{11}(\omega) = \lambda_{12}(\omega) = \lambda_{21}(\omega) = \lambda_{22}(\omega) = \lambda(\omega) = \sum_k \frac{|V_k|^2}{\omega - \omega_k}$, and $V_{12} = V_{21}$, $g(\omega) \equiv (\lambda(\omega) + V_{12})(\lambda(\omega) + V_{21}) - (\omega - \omega_1 - \lambda(\omega))(\omega - \omega_2 - \lambda(\omega))$ then Equation S3.3.6 can be simplified, where the $-\lambda_{22}(\omega)$ and $\lambda_{12}(\omega)$ terms are canceled out since $V_{k1}^* = V_{k2}^*$. The simplified Greens function is given by

$$D_{qk}(\omega) = \frac{\delta_{qk}}{\omega - \omega_q} - \frac{2V_q V_k^* \left(\omega - \frac{\omega_1 + \omega_2}{2} + V_{12}\right)}{(\omega - \omega_q)(\omega - \omega_k)g(\omega)}$$ (Equation S3.3.7)

Where the coefficients are defined as $g(\omega) \equiv (\lambda(\omega) + V_{12})^2 - (\omega - \omega_1 - \lambda(\omega))(\omega - \omega_2 - \lambda(\omega))$, and $\lambda(\omega) = \sum_k \frac{|V_k|^2}{\omega - \omega_k + i0^+}$. In the case where the two defects are equal with $\omega_1 = \omega_2 = \omega_0$, i.e. two-level degeneracy, the Equation S3.3.7 can further be simplified as

$$D_{qk}(\omega) = \frac{\delta_{qk}}{\omega - \omega_q} + \frac{2V_q V_k^*}{(\omega - \omega_q)(\omega - \omega_k)(\omega - \omega_0 - 2\lambda(\omega) - V_{12})}$$ (Equation S3.3.8)

To proceed, imagine there are many defects there, performing impurity average procedure, on average the phonon momentum will not change,[11] i.e., $\langle V_q V_k^* \rangle = n_i |V_k|^2 \delta_{qk}$, in which $n_i$ represents the density of the impurities, or equivalently filler volume fraction. For weak scattering regime, the impurity-averaged phonon propagator from Equation S3.3.7 can be written as

$$D_k(\omega) = \langle D_{qk}(\omega) \rangle \approx \frac{1}{\omega - \omega_k - \Sigma(k, \omega)}$$ (Equation S3.3.9)

Where the phonon self-energy correction $\Sigma(\mathbf{k}, \omega)$ can be written as

$$\Sigma(\mathbf{k}, \omega) = -\frac{2n_i |V_k|^2 \left(\omega - \frac{\omega_1 + \omega_2}{2} + V_{12}\right)}{g(\omega)}$$ (Equation S3.3.10)

For the degenerate defects $\omega_1 = \omega_2 = \omega_0$, which can be used to describe the perfect fillers (and defective fillers in an approximate way), we have

$$\Sigma(\mathbf{k}, \omega) = \frac{n_i |V_k|^2}{\frac{\omega - \omega_0 - V_{12}}{2} - \lambda(\omega)} \quad \text{(Equation S3.3.11)}$$

from which we can readily write down the real and imaginary parts as

$$Re\, \Sigma(\mathbf{k}, \omega) = \frac{n_i |V_k|^2 \left(\frac{\omega - \omega_0 - V_{12}}{2} - Re\, \lambda(\omega)\right)}{\left(\frac{\omega - \omega_0 - V_{12}}{2} - Re\, \lambda(\omega)\right)^2 + (Im\, \lambda(\omega))^2}$$

$$Im\, \Sigma(\mathbf{k}, \omega) = \frac{n_i |V_k|^2 Im\, \lambda(\omega)}{\left(\frac{\omega - \omega_0 - V_{12}}{2} - Re\, \lambda(\omega)\right)^2 + (Im\, \lambda(\omega))^2}$$

(Equation S3.3.12)

Where we have

$$Re\, \lambda(\omega) = \sum_k P \frac{|V_k|^2}{\omega - \omega_k} \quad \text{(Equation S3.3.13)}$$

$$Im\, \lambda(\omega) = -\pi \sum_q |V_q|^2 \delta(\omega - \omega_q) \quad \text{(Equation S3.3.14)}$$

### 3.4 Kubo formula for lattice thermal conductivity of defective polymers.

The generic thermal conductivity computed from the normal Green's function approach can be written as[12]

$$\kappa(T) = \frac{k_B \beta}{3L^3} \lim_{\delta \to 0} \int_0^{+\infty} e^{-\delta t} dt \int_0^{\beta} d\lambda \langle \mathbf{S}(0) \cdot \mathbf{S}(t + i\lambda) \rangle \quad \text{(Equation S3.4.1)}$$

Where the energy flow vector operator $S$ can be written as $\mathbf{S}(t) = \sum_k v_k \omega_k n_k(t)$, with $n_k = b_k^+ b_k$ is the phonon number density operator, $v_k$ is the phonon group velocity and $\omega_k$ is the dispersion. It has been shown that the phonon thermal conductivity Equation S3.4.1 can be rewritten in terms of the phonon Green's function as[13]

$$\kappa(T) = \frac{k_B \beta^2}{3\pi L^3} \sum_{kq} v_k \cdot v_q \omega_k \omega_q \times \int_{-\infty}^{+\infty} d\omega \frac{e^{+\beta \omega}}{(e^{\beta \omega} - 1)^2} Im\, D_{qk}(\omega) Im\, D_{kq}(\omega) \quad \text{(Equation S3.4.2)}$$

Where the phonon Green's function is defined as retarded form of Equation S3.2.1 to ensure consistency, and not other more common form of displacement-displacement correlator. In the case of phonon interacting with a two-level defects, $D_{qk}(\omega)$ can be expressed in terms of Equation S3.3.7.

As a sanity check, if we have phonon propagation written as

$$D_{qk}(\omega) = \frac{\delta_{qk}}{\omega - \omega_k + i\Gamma_k(\omega)} \quad \text{(Equation S3.4.3)}$$

Then we have

$$Im\, D_{qk}(\omega)\, Im\, D_{kq}(\omega) = \delta_{kq}\left(\frac{\Gamma_k(\omega)}{(\omega-\omega_k)^2+\Gamma_k^2(\omega)}\right)^2 \approx \frac{\pi\delta_{kq}\Gamma_k(\omega)\delta(\omega-\omega_k)}{(\omega-\omega_k)^2+\Gamma_k^2(\omega)},$$ and the total thermal conductivity Equation S3.4.2 can be written as

$$\kappa(T) = \frac{k_B\beta^2}{3L^3}\sum_k v_k^2 \omega_k^2 \frac{e^{+\beta\omega}}{(e^{\beta\omega}-1)^2}\frac{1}{\Gamma_k(\omega)} = \frac{1}{3}\sum_k v_k^2 \tau_k C_k \quad \text{(Equation S3.4.4)}$$

Where $\tau_k = \frac{1}{\Gamma_k}$ and $C_k = \frac{\omega_k \partial_T n_B(\omega_k)}{L^3}$ are phonon relaxation time and specific heat of a phonon with wavevector $\mathbf{k}$, $n_B(\omega_k)$ is the Bosonic occupation.

As to thermal diffusivity, we need the total specific heat capacity,

$$C(T) = \sum_k C_k(T) = \sum_k \frac{\omega_k \partial_T n_B(\omega_k)}{L^3} = \frac{1}{L^3}\sum_k \frac{\omega_k^2 e^{\beta\omega_k}}{T^2(e^{\beta\omega_k}-1)^2}$$
$$= \int d\omega\, \text{Dos}(\omega)\frac{\omega^2 e^{\beta\omega}}{T^2(e^{\beta\omega}-1)^2} \quad \text{(Equation S3.4.5)}$$

In all formula, we can consider the phonon energy $\omega_k$ we should use the renormalized phonon energy $\omega_k'$,

$$\omega_k' = \omega_k + Re\,\Sigma(\mathbf{k},\omega)$$
$$\Gamma_k(\omega) = -Im\,\Sigma(\mathbf{k},\omega) \quad \text{(Equation S3.4.6)}$$

Now, the overarching goal is clear, that we would like to explain the decrease of specific heat capacity $C(T)$ with respect to filler fraction $n_i$ using Equation S3.4.5, while the contradictory observation that the $\kappa(T)$ thermal conductivity actually increases using Equation S3.4.2. Normally, due to Equation S3.4.4, specific heat capacity and thermal conductivity trend are the same.

### 3.5 The connection of the experimental data.

The main goal to explain the data as the following, where $C(T)$ decreases vs $n_i$, while $K(T)$ increases vs $n_i$, even though from a kinetic model Equation S3.4.4 they are tightly linked to each other and often share the same trend. To see that, we notice that in the formula of $C(T)$

Equation S3.4.5, each term $C_k = \omega_k \partial_T n_B(\omega_k)$ is a monotonically decrease function of phonon energy $\omega_k$, and therefore the decrease of $C_k$ shall be linked to an increase of $\omega_k$. According to Equation S3.4.6, that means $Re\,\Sigma(\boldsymbol{k},\omega) > 0$.

On the other hand, after defect scattering, $\tau_k$ decreases, and according to Equation S3.4.4, the only possibility is an enhancement of phonon velocity $v_k$.

$$v_k = \partial_k \omega_k + \partial_k Re\,\Sigma(\boldsymbol{k},\omega) \quad \text{(Equation S3.5.1)}$$

i.e. a very large $\boldsymbol{k}$-variation of the reciprocal-space of the phonon self energy. Overall, by comparing with Equation S3.3.12, we anticipate that means $\frac{dV_k}{dk}$ large, which could mean a large heterogeneity that increases phonon group velocity and dominates over the scattering mechanism and the resulted reduction of lifetime and heat capacity. More quantitative calculations can be done but the contradicting behaviors of $C(T)$ and $K(T)$ constrain the possibilities:

$$\begin{aligned}\omega_k \uparrow, v_k \uparrow\uparrow,\\ \tau_\mathbf{k} \downarrow, C_\mathbf{k} \downarrow\downarrow,\\ C(T) \downarrow, K(T) \uparrow\end{aligned} \quad \text{(Equation S3.5.2)}$$

**3.6 Numerical analysis.**

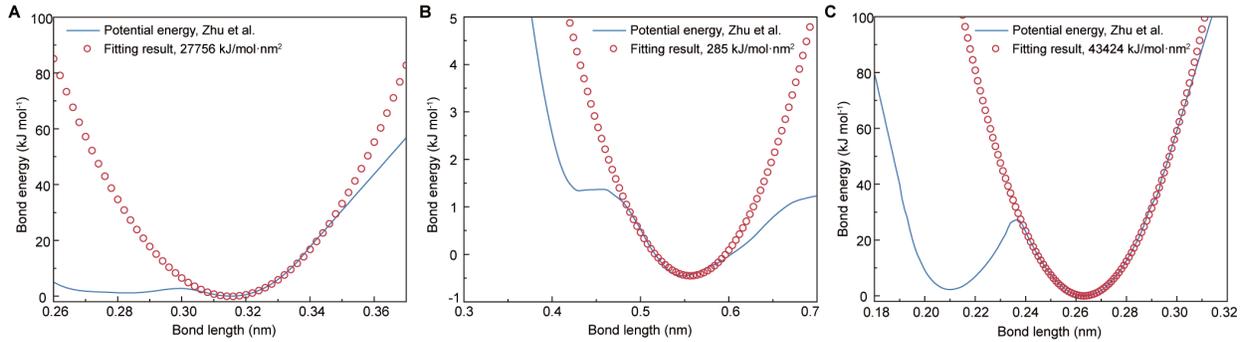

**Figure S16**. Relationships between coarse-grain potential to bond lengths in various materials.[14] (A) polymer-polymer (PVA-PVA). (B) polymer-filler (PVA-graphite oxide). (C) filler-filler (graphite oxide-graphite oxide).

To quantify the trend of $C(T)$, and $K(T)$ as a function of filler fraction $n_i$, we assume $\omega_k$ to take the form of the phonon dispersion of 1-D polymer chain. Furthermore, we consider $\omega_1$ is a constant, assuming a single vibrational state defect. The peak value of $\omega_k$ and $\omega_1$ is related to the square root of force constant between the polymer backbone and the fillers obtained by fitting quadratic function near the bond length between the beads. $V_k$ is assumed to take the form of Lorentzian function peaked at the center between Gamma point and the zone boundary, where the peak value of $V_k$ scale with the force constant between the polymer and filler. The numerical analysis is performed using the parameters in Table S1. Finally, it is also observed that the heat capacity $C(T)$ with defective fillers is consistently lower than that of the perfect fillers, while the thermal conductivity $K(T)$ of polymer composites with defective fillers is consistently higher (Figure 2D and 2F, respectively). This can be readily explained through the additional vibrational defects level, by assuming $\omega_1 \neq \omega_2$ in Equation S3.3.11. Under the condition that the effective phonon group velocity increases, shown schematically in Equation S3.5.2, then the observation can be fully reproduced. Physically, this can be understood since more defect levels couple with polymers with more channels, and effectively hardens the polymer.

**Table S1**. The value of parameters used in the numerical analysis.

| | |
|---|---|
| $\omega_k$ | 5.07e-21 J |
| $\dfrac{\omega_1}{\omega_k}$ | [0.08, 0.5] |
| $\dfrac{V_k}{\omega_k}$ | [0.08, 1.4] |
| $k_b T$ | 4.02e-21 J |
| T | 290 K |
| $V_{12}$ | 0 |
| $0^+$ | 0.1 * \omega |
| $n_i$ | [25, 50] |

To quantify the trend of $C(T)$, and $K(T)$ as a function of filler fraction $n_i$, we assume $\omega_k$ to take the form of the phonon dispersion of 1-D polymer chain. Furthermore, we consider $\omega_1$ is a constant, assuming a single vibrational state defect. The peak value of $\omega_k$ and $\omega_1$ is related to the square root of force constant between the polymer backbone and the fillers obtained by fitting quadratic function near the bond length between the beads. $V_k$ is assumed to take the form of Lorentzian function peaked at the center between Gamma point and the zone boundary, where the peak value of $V_k$ scale with the force constant between the polymer and filler. The numerical analysis is performed using the parameters in Table S1.

Finally, it is also observed that the heat capacity $C(T)$ with defective fillers is consistently lower than that of the perfect fillers, while the thermal conductivity $K(T)$ of polymer composites with defective fillers is consistently higher. This can be readily explained through the additional vibrational defects level, by assuming $\omega_1 \neq \omega_2$ in Equation S3.3.11. Under the condition that the effective phonon group velocity increases, shown schematically in Equation S3.5.2, then the observation can be fully reproduced. Physically, this can be understood since more defect levels couple with polymers with more channels, and effectively hardens the polymer. The ratio $\frac{\omega_1}{\omega_k}$ and $\frac{V_k}{\omega_k}$ from the force constant in Figure S16 are approximately 1.25 and 0.1, respectively which shows the decreasing trend of $C(T)$ and increasing trend of $K(T)$ as observed in the experiment.